\documentclass[12pt]{article}
\pdfoutput=1
\usepackage{latexsym}
\usepackage{amsmath,,calrsfs}
\usepackage{amsfonts}
\usepackage{amssymb}
\usepackage{amscd}
\usepackage{bbm}
\usepackage{fancybox}
\usepackage{cite}
\usepackage{amsmath,amsfonts,amsbsy}
\usepackage{pstricks,pst-node}
\usepackage[small,bf,hang]{caption2}
\usepackage{graphicx}
\usepackage{epsfig}
\usepackage{psfrag}
\usepackage{comment}

\usepackage{hyperref}

\usepackage{float}

\psset{unit=1.3cm,linewidth=.5pt,radius=.2}  

\usepackage{multirow}                     
\usepackage{float}                          
\usepackage{lscape}                         
\usepackage{bm}

\newcommand{\cN}{{\cal N}}

\newcommand{\bi}{\begin{itemize}}
\newcommand{\ei}{\end{itemize}}
\newcommand{\bea}{\begin{eqnarray}}
\newcommand{\eea}{\end{eqnarray}}
\newcommand{\bt}{\begin{tabular}}
\newcommand{\et}{\end{tabular}}
\newcommand{\bc}{\begin{center}}
\newcommand{\ec}{\end{center}}

\def\theequation{\arabic{section}.\arabic{equation}}

\newcommand{\be}{\begin{equation}}
\newcommand{\ee}{\end{equation}}
\newcommand{\ba}{\begin{array}}
\newcommand{\ea}{\end{array}}
\newcommand{\p}[1]{(\ref{#1})}
\newcommand{\lb}[1]{\label{#1}}

\def\bbox{{\,\lower0.9pt\vbox{\hrule \hbox{\vrule height 0.2 cm
\hskip 0.2 cm \vrule height 0.2 cm}\hrule}\,}}
\newcommand{\dsl}{\pa \kern-0.5em /}

\newcommand{\nn}{\nonumber \\}

\makeatletter \@addtoreset{equation}{section} \makeatother
\renewcommand{\theequation}{\thesection.\arabic{equation}}

\def\slashchar#1{\setbox0=\hbox{$#1$}           
   \dimen0=\wd0                                 
   \setbox1=\hbox{/} \dimen1=\wd1               
   \ifdim\dimen0>\dimen1                        
      \rlap{\hbox to \dimen0{\hfil/\hfil}}      
      #1                                        
   \else                                        
      \rlap{\hbox to \dimen1{\hfil$#1$\hfil}}   
      /                                         
   \fi}

\begin{document}

\begin{titlepage}

\renewcommand{\thefootnote}{\star}

\begin{center}

\hfill  {}


{\Large \bf  Unconstrained off-shell superfield formulation
 \vspace{0.2cm}

of $4D, \mathcal{N}=2$ supersymmetric higher spins}

\vspace{1cm}
\renewcommand{\thefootnote}{$\star$}

{\large\bf Ioseph Buchbinder} ${}^{\dag, \, +}$, {\quad \large\bf
Evgeny~Ivanov} ${}^{\ast,\, \star}$, {\quad \large\bf
Nikita~Zaigraev} ${}^{\ast,\, \star}$
 \vspace{1.3cm}


{${}^\dag$ \it Center of Theoretical Physics, Tomsk State Pedagogical University,} \\
{\it 634061, Tomsk,  Russia} \\

\vskip 0.15cm

{ ${}^+$ \it National Research Tomsk State University, 634050, Tomsk, Russia} \\
\vskip 0.15cm

{${}^\ast$ \it Bogoliubov Laboratory of Theoretical Physics, JINR,}\\
{\it 141980 Dubna, Moscow region, Russia} \\
\vskip 0.15cm

{${}^\star$ \it Moscow Institute of Physics and Technology,}\\
{\it 141700 Dolgoprudny, Moscow region, Russia}
\vspace{0.2cm}

{\tt joseph@tspu.edu.ru, eivanov@theor.jinr.ru, nikita.zaigraev@phystech.edu}\\

\end{center}
\vskip 0.6truecm \nopagebreak

\begin{abstract}
\noindent We present, for the first time, the complete off-shell
$4D, {\cal N}=2$ superfield actions for any free massless integer
spin ${\bf s} \geq 2$ fields, using the ${\cal N}=2$ harmonic
superspace approach. The relevant gauge supermultiplet is
accommodated by two real analytic bosonic superfields
$h^{++}_{\alpha(s-1)\dot\alpha(s-1)}$,
$h^{++}_{\alpha(s-2)\dot\alpha(s-2)}$ and two conjugated complex
analytic spinor superfields $h^{+3}_{\alpha(s-1)\dot\alpha(s-2)}$,
$h^{+3}_{\alpha(s-2)\dot\alpha(s-1)}\,$, where $\alpha(s) := (\alpha_1\ldots \alpha_s),
\dot\alpha(s) := (\dot\alpha_1\ldots \dot\alpha_s)$. Like in the harmonic
superspace formulations of ${\cal N}=2$ Maxwell and supergravity
theories, an infinite number of original off-shell degrees of
freedom is reduced to the finite set (in WZ-type gauge) due to an
infinite number of the component gauge parameters in the analytic
superfield parameters. On shell, the standard spin content $({\bf s,
s-1/2, s-1/2, s-1})$ is restored. For ${\bf s}=2$ the action
describes the linearized version of ``minimal'' ${\cal N}=2$
Einstein supergravity.
\end{abstract}
\vskip 1cm

\begin{center}
{\it Dedicated to Emery Sokatchev on the occasion of his 70th birthday}
\end{center}

\vspace{1cm}

\phantom{Keywords: }
\vspace{1cm}

\newpage

\end{titlepage}

\setcounter{footnote}{0}

\newpage
\setcounter{page}{1}

\section{Introduction}
Supersymmetric higher-spin field theories attract a vast attention during a long time.
There are at least two reasons for this. First, from the purely theoretical point of view,
it is tempting to construct new supersymmetric models of this kind, as well as to supersymmetrize
the already available higher-spin field bosonic models. The new universal methods to be developed
during these studies could, in particular, shed more light on hidden relationships between fermionic and bosonic degrees of
freedom for the higher-spin fields and open  new
possibilities for building consistent higher-spin field models due
to the appearance of extended gauge (super)symmetries. Second, since the superstring theory
encompasses infinite towers of bosonic and fermionic higher spin states,
supersymmetric higher-spin gauge theory can serve as a bridge between
superstring theory and low-energy field theory.

There is a huge literature on higher spin fields. In this
introductory  section (and over the whole work) we limit our
discussion to the issues related {\it only to supersymmetric
higher spin theories} and {\it only to the four-dimensional}
versions of the latter. Respectively, our reference list mainly
includes the papers of the same trend.

As is well known, there exist two different generic formulations of
the supersymmetric field theories, the component (on-shell)
formulation and superfield (off-shell) formulation (see, e.g.,
\cite{GGRS}, \cite{BL}, \cite{18}). In the first approach, the
theory is formulated in a way lacking a manifest  supersymmetry, in
terms of bosonic and fermionic fields forming a supermultiplet on
the mass shell. The algebra of the supersymmetry transformations is
open and becomes closed only upon using the equations of motion. To
close the algebra off shell, we are led to introduce the auxiliary
fields which vanish on the mass shell. In the second approach, a
theory is formulated in a manifestly supersymmetric way employing
superfields. The  algebra of the supersymmetry transformations is
automatically off-shell closed and the auxiliary fields are already
built in the superfields. Clearly, due to the manifest off-shell
supersymmetry, the second approach much more suits for setting up
various generalizations, e.g., finding out consistent interactions
proceeding from a given free theory. However, the superfield
formulations of some Lagrangian field theory (both classical and
quantum) are useful and efficient only providing that the relevant
superfields are not subject to any algebraic constraints. For all
types of supersymmetries in all dimensions, such unconstrained
superfield formulations are at present generically unknown. What
concerns four dimensions, there exist unconstrained superfield
formulations for all ${\cal N}=1$ supersymmetric theories of
interest (matter, super Yang-Mills, supergravity)  in conventional
$4D$ superspaces, general and chiral (see, e.g., \cite{GGRS},
\cite{BL}), and for their  ${\cal N}=2$ supersymmetry counterparts
in terms of harmonic superspace \cite{HSS,18}.

Free massless bosonic and fermionic higher-spin field theories have
been pioneered by Fronsdal \cite{FronsdalInteg},
\cite{FronsdalHalfint}. The corresponding supersymmetric
generalization in the component approach can be constructed as
follows. Lagrangian is written as a sum of Lagrangians for all
fields of the on-shell supermultiplet. Then one should invent the
appropriate supersymmetry transformations and check the invariance
of the total Lagrangian. Such a description was realized for $4D$
free massless higher-spin ${\cal N}=1$ supersymmetric models in
works \cite{Courtright}, \cite{Vasiliev} \footnote{Later, it was
shown that the free supersymmetric massless higher-spin gauge theory can
be also formulated in the framework of the BRST formalism
\cite{BKoutr}.}. Complete off-shell Lagrangian formulation of $4D$
free higher-spin ${\cal N}=1$ models has been developed in terms of
${\cal N}=1$ superfields in works \cite{Kuz1}, \cite{Kuz2},
\cite{Kuz3} (see also section 6.9 in \cite{BL}) and further applied
to study quantum effective action generated by ${\cal N}=1$
superfields in ${\rm AdS}$ space in \cite{BKS}. Note that in \cite{Kuz3} the massless higher-spin ${\cal N}=1$ supermultiplets in AdS$_4$
were constructed for the first time. The superfield approach to
${\cal N} = 1$ supersymmetric massless higher spin fields was
further generalized in \cite{GKS1}, \cite{GKS2}, \cite{GKuz},
\cite{Kuzenko:2017ujh}, \cite {HK1}, \cite{HK2}, \cite{BHK},\cite{Buchbinder:2019yhl}
\cite{KPR}\footnote{It is worth noting the paper
\cite{Kuzenko:2021pqm}, where $4D, {\cal N}=2$ superconformal higher-spin theory
was formulated in terms of unconstrained ${\cal N}=2$ superfields.
The program of constructing the massless higher spin $\cN=2$
superfield actions was sketched but not realized there. Here we do
not deal with the superconformal theories at all.}. Some additional
geometric aspects of this approach were explored in \cite{GK1},
\cite{GK2}, \cite{BGK1}, \cite{BGK2}. Also note an activity on
${\cal N}=1$ supersymmetric massive higher spin theories (see e.g.,
\cite{Z1}, \cite{Z2}, \cite{K}  and the references therein), however
it is out of the subject of our paper.

At present, a manifestly supersymmetric off-shell
unconstrained superfield Lagrangian formulation for extended
higher-spin supersymmetric theories is unknown even for the free
case (modulo the superconformal theories \cite{Kuzenko:2021pqm} which we do not concern here). Progress in this area is associated either with the
realization of extended supersymmetry in terms of ${\cal N}=1$
superfields \cite{GKS1}, \cite{GKS2}, or in terms of light-cone ${\cal N}$- extended
superfields \cite{M2}, or in the on-shell component approach (see,
e.g., \cite{BS} and the references therein). In all cases, the full
extended supersymmetry remains non-manifest. An off-sell Lagrangian
formulation of $\cN=2$ supersymmetric higher-spin theory on AdS
space in terms of unconstrained  $\cN=1$ superfields (and some
Poincar\'e supersymmetric limits thereof) has been constructed for the first time in \cite{GKS2} but
such a formulation does not reveal a manifest $\cN=2$ supersymmetry. As a result we
can conclude that the problem of complete off-shell description of
the higher-spin extended supersymmetric theories is still open
\footnote{The free equations of motion for higher-spin massless
${\cal N}=2$ superfields have been constructed in the conventional
${\cal N}=2$ AdS superspace in ref. \cite{SeSi}. As noted by its
authors, the issue of constructing the corresponding Lagrangian
formulation remained unresolved in their approach.}.

Note that one of the actively developing directions in the theory of
higher spin fields is related with the study of interactions. In
particular, recently a substantial understanding of the structure of
cubic interaction vertices for higher spin supersymmetric fields has
been achieved in different component and ${\cal N}=1$ superfield
approaches (see, e.g., \cite{M2}, \cite{M1}, \cite{KhZ}, \cite{BKTW},
\cite{BuchGK3}, \cite{BuchGK4}, \cite{GaK} and the references
therein).
Other aspects of higher-spin supersymmetric field theory are related with
supersymmetric  extension \cite{Sezgin1},
\cite{Engquist:2002vr},\cite{Sezgin2} of the Vasiliev theory of
interacting higher spin fields (see the reviews \cite{V1},
\cite{V2}, \cite{V3},
and the references therein). It is beyond the scope of our paper to
discuss these prospective and advanced studies.

In this paper we construct the completely off-shell manifestly
${\cal N}=2$ supersymmetric superfield extension of arbitrary $4D$
integer-spin free massless theory. The construction is based on the
use of the harmonic superspace method \cite{18} which is at present
the most adequate and convenient approach for description of $4D,
{\cal N}=2$ supersymmetric field theories.

The paper is organized as follows. Section 2 is devoted to a brief
description of the linearized ${\cal N}=2$ Einstein supergravity
(linearized massless ${\cal N}=2$ spin 2 theory) in terms of
unconstrained analytic harmonic superfields. In section 3 we generalize the
above results and formulate the free massless ${\cal N}=2$ spin 3
harmonic superfield theory. Section 4 is devoted to further
generalization and construction of a completely off-shell invariant
action for the free massless ${\cal N}=2$ gauge theory with
an arbitrary maximal integer spin ${\bf s}$ of the supermultiplet. The theory
is formulated in  $4D, {\cal N}=2$ harmonic superspace in terms of
unconstrained analytic superfields. In section 5 we summarize the results and
discuss possible ways of further development of the approach presented.

\section{$\mathcal{N}=2$ spin 2 theory}

\subsection{Minimal Einstein $\mathcal{N}=2$ supergravity in the harmonic approach}

We start by a sketch of the basic principles of Einstein
$\mathcal{N}=2$ supergravity (SG) in harmonic superspace
\cite{Galperin:1987em,18}. Its linearized version provides an
off-shell ${\cal N}=2$ supersymmetric free spin 2 action and will
serve as a prototype for constructing ${\cal N}=2$ higher spin
actions.

We will deal with ${\cal N}=2$ harmonic superspace (HSS) in the analytic
basis as the following set of coordinates \cite{HSS,18}
\be
 Z =
\big(x^m, \theta^{+\mu}, \bar\theta^{+\dot\mu}, u^\pm_i,
\theta^{-\mu}, \bar\theta^{-\dot\mu}\big) \equiv \big(\zeta,
\theta^{-\mu}, \bar\theta^{-\dot\mu}\big), \lb{HSS}
\ee
where the
standard notation of ref. \cite{18} is used. In particular,
$u^\pm_i$ are harmonic variables parametrizing the internal sphere
$S^2$, $u^{+i}u^-_i =1$, the indices $\pm$ denote the harmonic
$U(1)$ charges of various quantities and the index $i = 1, 2$ is the
doublet index of the automorphism $SU(2)$ group acting only on the
harmonic variables. The set \p{HSS} is closed under the rigid ${\cal
N}=2$ supersymmetry transformations
\be
\delta_\epsilon x^m = -2i
\big(\epsilon^-\sigma^m \bar\theta^+ + \theta^+\sigma^m
\bar\epsilon^-\big), \; \delta_\epsilon\theta^{\pm \hat \mu} =
\epsilon^{\pm \hat \mu}\,, \; \delta_\epsilon u^{\pm}_i = 0\,, \quad
\epsilon^{\pm \hat \mu} = \epsilon^{\hat \mu i } u^\pm_i\,,
\lb{N2SUSY}
\ee
where we employed the condense notation, $\hat\mu =
(\mu, \dot\mu)$. These transformations also leave intact the
harmonic analytic subspace of \p{HSS},
\be
\zeta := \big(x^m,
\theta^{+\mu}, \bar\theta^{+\dot\mu}, u^\pm_i\big). \lb{AHSS}
\ee
The HSS formulation of ${\cal N}=2$ SG is displayed in an extension
of the HSS \p{HSS} by a fifth coordinate $x^5$,
\be
Z \Longrightarrow (Z, x^5)\,,\lb{Ext5}
\ee
with the following
analyticity-preserving transformation law under ${\cal N}=2$
supersymmetry,
\be
\delta_\epsilon x^5 = 2i\big(\epsilon^-\theta^+ -
\bar\epsilon^-\bar\theta^+ \big). \lb{Tranfifth}
\ee
This coordinate
can be interpreted as associated with the central charge in ${\cal N}=2$ Poincar\'e superalgebra.

An important ingredient of the HSS formalism is the harmonic
derivatives $D^{++}$ and $D^{--}$ which have the following form in
the analytic basis \footnote{Hereafter, we use the notations
$\hat{\mu} \equiv (\mu, \dot{\mu})$, $\partial^\pm_{\hat{\mu}} =
\partial/ \partial \theta^{\mp \hat{\mu}}$, $(\theta^{\hat{+}})^2
\equiv (\theta^+)^2 - (\bar{\theta}^+)^2$ and
$\partial_{\alpha\dot\alpha} =
\sigma^m_{\alpha\dot\alpha}\partial_m$. The summation rules are
$\psi\chi = \psi^\alpha \chi_\alpha, \bar\psi\bar\chi =
\bar\psi_{\dot\alpha} \bar\chi^{\dot\alpha}$, Minkowski metric is
${\rm diag} (1, -1, -1, -1)$ and $\Box = \partial^m\partial_m =
\frac12 \partial^{\alpha\dot\alpha}\partial_{\alpha\dot\alpha}$.}
\begin{eqnarray}
    && {D}^{++} = \partial^{++} - 2i \theta^{+\rho} \bar{\theta}^{+\dot{\rho}} \partial_{\rho\dot{\rho}} + \theta^{+\hat{\mu}} \partial^{+}_{\hat{\mu}}
    +
    i (\theta^{\hat{+}})^2 \partial_5\,, \nonumber \\
&& D^{--} = \partial^{--}- 2i \theta^{-\rho} \bar{\theta}^{-\dot{\rho}} \partial_{\rho\dot{\rho}} + \theta^{-\hat{\mu}} \partial^{-}_{\hat{\mu}}
    +
    i (\theta^{\hat{-}})^2 \partial_5\,, \lb{Dflat} \\
&& [D^{++}, D^{--}] = D^0\,, \quad D^0 = u^{+ i}\frac{\partial}{\partial u^{+i}} - u^{- i}\frac{\partial}{\partial u^{-i}} + \theta^{+ \hat\mu}\partial^-_{\hat\mu}
-\theta^{- \hat\mu}\partial^+_{\hat\mu}\,.\lb{Flatness}
\end{eqnarray}
The crucial difference between derivatives $D^{++}$ and $D^{--}$ is that $D^{++}$ preserves analyticity,
while $D^{--}$ does not.

We will be interested in the simplest version of Einstein ${\cal N}=2$ SG which is obtained from the conformal ${\cal N}=2$ SG by invoking
the so called nonlinear multiplet as one of the two necessary compensating multiplets. In the HSS formalism, one uses a gauge in which the analytic
superfield which accommodates this
compensating multiplet is gauged away to yield the fundamental group of the resulting Einstein ${\cal N}=2$ SG as the following
analyticity-preserving superdiffeomorphisms
\begin{eqnarray}
&& \delta_\lambda x^m = \lambda^m (x, \theta^+, u), \quad \delta_\lambda x^5 = \lambda^5 (x, \theta^+, u)\,, \lb{xx5} \\
&& \delta_\lambda \theta^{+\mu} = \lambda^{+\mu} (x, \theta^+, u),
    \qquad
    \delta_\lambda \bar{\theta}^{+\dot{\mu}} = \bar{\lambda}^{+\dot{\mu}} (x, \theta^+, u), \nn
&&    \delta_\lambda \theta^{-\mu} = \lambda^{-\mu} (x, \theta^+, \theta^-, u),
    \qquad
    \delta_\lambda \bar{\theta}^{-\dot{\mu}} = \bar{\lambda}^{-\dot{\mu}} (x, \theta^+, \theta^-, u), \lb{N2Diff} \\
&& \delta_\lambda u_i^\pm =0.
\end{eqnarray}
Notice that neither the gauge parameters nor any of the geometrical objects used in the paper depend on the fifth coordinate $x^5$.

Next one defines a generalization of the flat harmonic derivatives \p{Dflat}, $\mathfrak{D}^{\pm\pm}$, such that they were covariant under \p{N2Diff}
\be
\delta \mathfrak{D}^{\pm\pm} = 0\; \Rightarrow{\;} \lb{CondInv}
\ee
\bea
&& \mathfrak{D}^{++} = D^{++} + h^{++m} \partial_m + h^{++\hat{\mu}+} \partial^{-}_{\hat{\mu}} + h^{++\hat{\mu}-} \partial^{+}_{\hat{\mu}}
+ h^{++5} \partial_5\,, \lb{covD++}\\
&& \mathfrak{D}^{--} = {D}^{--} + h^{--m} \partial_m + h^{--\hat{\mu}+} \partial^{-}_{\hat{\mu}} + h^{--\hat{\mu}-} \partial^{+}_{\hat{\mu}}
    +
   h^{--5} \partial_5\,.\lb{covD--}
\eea
The components of the vielbein $h^{++M}$ in \p{covD++} are analytic superfields, $h^{++M} = h^{++M}(\zeta)$, as the constraints of ${\cal N}=2$ SG in the HSS formulation
require that \cite{Galperin:1987em}
\begin{equation}\label{analytic}
    [\partial^+_{\hat{\mu}}, \mathfrak{D}^{++}] = 0\,.
\end{equation}
The negatively charged vielbeins in \p{covD--} are expressed in terms of those in \p{covD++} from the conditions implied by the harmonic constraint
\be
[\mathfrak{D}^{++}, \mathfrak{D}^{--}] = D^0\,, \lb{Flat2}
\ee
which is just a generalization of the flat superspace condition \p{Flatness}. The explicit form of the relevant relations will be given below for the linearized theory.
The transformation properties of the vielbeins $h^{++M}$ and $h^{--M}$ are uniquely determined by  \p{CondInv}, whence, in particular, $\delta_\lambda D^0 = 0$.
Here we present them for $h^{++M}$, postponing those for $ h^{--M}$ also until the linearized case,
\begin{eqnarray}\label{transf}
        &&\delta_\lambda h^{++m } = \mathfrak{D}^{++} \lambda^m + 2i \lambda^{+\alpha} \sigma^m_{\alpha\dot{\alpha}} \bar{\theta}^{+\dot{\alpha}}
        + 2i \theta^{+\alpha} \sigma^m_{\alpha\dot{\alpha}} \bar{\lambda}^{+\dot{\alpha}}\,,\nn
       && \delta_\lambda h^{++5} = \mathfrak{D}^{++} \lambda^5 - 2i  \lambda^{+\hat{\mu}} \theta^{+}_{\hat{\mu}} \,,\nn
       && \delta_\lambda h^{++\hat{\mu}+} = \mathfrak{D}^{++} \lambda^{+\hat{\mu}}\,, \nn
       && \delta_\lambda h^{++\hat{\mu}-} = \mathfrak{D}^{++} \lambda^{-\hat{\mu}}
        - \lambda^{+\hat{\mu}} \;.
\end{eqnarray}
Note, that the non-analytic vielbein $h^{++\hat{\mu}-}$ and the non-analytic parameter $\lambda^{-\hat{\mu}}$ have exactly the same component contents.
Therefore, this vielbein can be entirely gauged away,
\be
h^{++\hat{\mu}-} = 0\,.
\ee
In this gauge we have the ``analytic gauge'' condition

\begin{equation}\label{analytic gauge}
 \mathfrak{D}^{++}\lambda^{-\hat{\mu}} = \lambda^{+\hat{\mu}}\,,
\end{equation}
which fully specifies $\lambda^{-\hat{\mu}}$ in terms of the components of $\lambda^{+\hat{\mu}}$.

Now, using the transformations \p{transf}, one can display the field content of the vielbeins in the Wess-Zumino gauge:

\begin{eqnarray}\label{wesszumino-grav}
&&        h^{++m}
       =
        -2i \theta^+\sigma^a \bar{\theta}^+ \Phi^m_a
       +  (\bar{\theta}^+)^2 \theta^+ \psi^{m\,i}u^-_i + (\theta^+)^2 \bar{\theta}^+ \bar{\psi}^{m\,i}u_i^- +  (\theta^+)^2 (\bar{\theta}^+)^2 V^{m(ij)}u^-_iu^-_j,\nn
&&       h^{++5} =
       -2 i \theta^+ \sigma^a \bar{\theta}^+ C_a
        + (\bar{\theta}^+)^2 \theta^+ \rho^{i}u^-_i + (\theta^+)^2 \bar{\theta^+} \bar{\rho}^{i}u_i^- + (\theta^+)^2 (\bar{\theta}^+)^2 S^{(ij)}u^-_iu^-_j, \nn
&&       h^{++\mu+} = (\theta^+)^2 \bar{\theta}^+_{\dot{\mu}} P^{\mu\dot\mu}
       +  \left(\bar{\theta}^+\right)^2 \theta^+_\nu \left[\varepsilon^{\mu\nu}M + T^{(\mu\nu)}\right] \nn
&&      \;\;\;\;\;\;\;\;\;\;\; + \, (\theta^+)^2 (\bar{\theta}^+)^2 \chi^{\mu i}u^-_i\,, \qquad h^{++\dot{\mu}+} = \widetilde{h^{++\mu+}}\,.
\end{eqnarray}
This is just the content of the ``minimal'' $\mathcal{N}=2$ Einstein supergravity multiplet \cite{Fradkin:1979cw} (note that the fields $M, P^{\mu\dot\mu}, T^{(\mu\nu)}$ entering
$h^{++\mu+}$ in \p{wesszumino-grav} are complex). So the analytic superfields $h^{++m}, h^{++\hat{\mu}+}, h^{++5}$
are the unconstrained gauge potentials of  the ``minimal'' $\mathcal{N}=2$ Einstein supergravity. The physical fields are $\Phi^m_a, \psi^{m\,i}_{\hat\mu}, C_a$,
the remaining ones are auxiliary. After eliminating them from the appropriate action, we are left with the on-shell superspin 1, superisospin 0  multiplet $({\bf 2, 3/2, 3/2, 1})$.
With taking into account the residual gauge freedom of the WZ gauge \p{wesszumino-grav} (see below), the complete set of essential off-shell degrees of freedom  is ${\bf 40} + {\bf 40}$.

\subsection{Linearized theory}

In what follows, we will be interested in the linearized version of the above construction.

In general, the negatively charged vielbeins in \p{covD--} obey rather complicated nonlinear harmonic equations following from the condition \p{Flat2}.
 However, at the linearized level these conditions are
essentially simplified: they are reduced to the linear harmonic equations for the gauge potentials:
\bea
&& D^{++}h^{--\alpha\dot\alpha} - D^{--}h^{++\alpha\dot\alpha}  + 4i\big( h^{--\alpha+}\bar\theta^{+\dot\alpha} + \theta^{+\alpha}h^{--\dot\alpha+}\big) = 0\,, \nn
&&D^{++}h^{--5} - D^{--}h^{++5} -2i\big( h^{--\alpha+}\theta^{+}_\alpha - \bar\theta^+_{\dot\alpha}h^{--\dot\alpha+} \big) = 0\,, \lb{m5} \\
&& D^{++}  h^{--\alpha+} - D^{--} h^{++\alpha+} =0\,, \quad D^{++}  h^{--\dot\alpha+} - D^{--} h^{++\dot\alpha+} =0\,, \nn
&& D^{++}  h^{--\alpha-} - h^{--\alpha+} =0\,, \quad D^{++}  h^{--\dot\alpha-} - h^{--\dot\alpha+} =0\,. \lb{spin}
\eea
These constraints are invariant under the following linearized form of the superfield gauge transformations \p{transf}
and their counterparts for the negatively charged vielbeins
\bea
&&\delta_\lambda h^{++m } = {D}^{++} \lambda^m + 2i \big( \lambda^{+\alpha} \sigma^m_{\alpha\dot{\alpha}} \bar{\theta}^{+\dot{\alpha}}
        + \theta^{+\alpha} \sigma^m_{\alpha\dot{\alpha}} \bar{\lambda}^{+\dot{\alpha}}\big)\,, \nn
&&\delta_\lambda h^{++5} = {D}^{++} \lambda^5 - 2i \big(\lambda^{+{\alpha}} \theta^{+}_{\alpha} - \bar\theta^{+}_{\dot{\alpha}}\bar\lambda^{+\dot{\alpha}}\big), \nn
        && \delta_\lambda h^{++\hat{\mu}+} = {D}^{++} \lambda^{+\hat{\mu}}\,,  \lb{TranLin+} \\
&& \delta_\lambda h^{--m } = {D}^{--} \lambda^m + 2i \big(\lambda^{-\alpha} \sigma^m_{\alpha\dot{\alpha}} \bar{\theta}^{-\dot{\alpha}}
+ \theta^{-\alpha} \sigma^m_{\alpha\dot{\alpha}} \bar{\lambda}^{-\dot{\alpha}}\big)\,, \nonumber \\
&&\delta_\lambda h^{--5} = {D}^{--} \lambda^5 - 2i \big(\lambda^{-{\alpha}} \theta^{-}_{\alpha} - \bar\theta^{-}_{\dot{\alpha}}\bar\lambda^{-\dot{\alpha}}\big), \nn
&& \delta_\lambda h^{--{\mu}+} = {D}^{--} \lambda^{+{\mu}} - \lambda^{-\mu}\,, \quad \delta_\lambda h^{--\dot{\mu}+} =
{D}^{--} \lambda^{+\dot{\mu}} - \bar\lambda^{-\dot\mu}\,, \lb{TranLin-}\\
&& \delta_\lambda h^{--{\mu}-} = {D}^{--} \lambda^{-{\mu}}\,, \quad \delta_\lambda h^{--\dot{\mu}-}= {D}^{--} \lambda^{-\dot{\mu}}\,. \lb{TranLin-3}
\eea
These transformations can still be used to choose the Wess-Zumino gauge \p{wesszumino-grav} for analytic superfields $h^{++ M}$ in the linearized theory
as well, though in this approximation $h^{++ M}$ and $h^{-- M}$ loose their geometric meaning of vielbeins. Similarly, the analytic gauge parameters $\lambda^{++ m, 5}$ and
$\lambda^{+\hat\mu}$ loose
their original geometric meaning of the parameters of the coordinate superdiffeomorphisms preserving the analytic subspace \p{AHSS}. The non-analytic gauge parameter $\lambda^{-\hat\mu}$
satisfies the linearized form of eq. \p{analytic gauge},
\be
D^{++}\lambda^{-\hat\mu} = \lambda^{+\hat\mu}\,.
\ee

As usual, fixing WZ gauge does not fully capture symmetry. The residual gauge freedom of the theory is spanned by the parameters:
\begin{equation}
    \begin{cases}
        \lambda^m \;\Rightarrow\; a^m(x)\,, \\
        \lambda^5 \; \Rightarrow \; b(x)\,,  \\
        \lambda^{\mu+} \;\Rightarrow \; \epsilon^{\mu i}(x) u^+_i + \theta^{+\nu}l_{(\nu}^{\;\;\;\mu)}(x)\,, \\
        \bar{\lambda}^{\dot{\mu}+} \;\Rightarrow \; \bar{\epsilon}^{\dot{\mu} i}(x) u^+_i + \bar{\theta}^{+\dot{\nu}}l_{(\dot{\nu}}^{\;\;\;\dot{\mu})}(x)\,. \lb{WZsymm2}
    \end{cases}
\end{equation}
It is natural to make the following identification:

\begin{itemize}
    \item $a^m(x)$ are the remnants of the diffeomorphism parameters which  now form the basic gauge freedom of the free spin 2 field;
    \item $b(x)$ is the parameter of Abelian gauge transformations acting on the ``graviphoton'' $A^m$;
    \item $\epsilon^{\hat{\mu}i}(x)$ originate from the parameters of local supersymmetry which are now ${\cal N}=2$ counterparts
    of the local $a^m$ transformations;
    \item $l^{(\mu\nu)}$ and $l^{(\dot{\mu}\dot{\nu})}$ are the former parameters of local Lorentz transformations which can be used to gauge away
    the antisymmetric part of $\Phi^m_a$ and so to leave  only the symmetric part in the latter (traceless ``conformal graviton'' and the trace itself).
\end{itemize}

For the further consideration, it will be instructive to explicitly see (before any gauge-fixing)  how the gauge freedom \p{TranLin+} allows to remove
 all the ``superfluous'' $SU(2)$ singlet bosonic spins from the basic gauge superfields. The relevant shifting local symmetries are contained in the supergauge parameter
$\lambda^{+\alpha}, \bar\lambda^{+\dot\alpha}$, while the physical dimension $SU(2)$ singlet spins in $h^{++m,5}$. Passing, for the convenience, to the spinor notation,
$h^{++}_{\alpha\dot\alpha} = (\sigma_m)_{\alpha\dot\alpha} h^{++ m}\,,$ we identify these particular components as
\bea
&& h^{++}_{\alpha\dot\alpha} \;\Rightarrow \; (\theta^+)^2 \omega_{\alpha\dot\alpha} + (\bar\theta^+)^2 \bar\omega_{\alpha\dot\alpha}
-2i \theta^{+\beta}\bar\theta^{+\dot\beta}\Phi_{\beta\dot\beta \alpha\dot\alpha}\,, \nn
&& h^{++ 5} \;\Rightarrow \; (\theta^+)^2 \omega +
(\bar\theta^+)^2 \bar\omega
+i \theta^{+\beta}\bar\theta^{+\dot\beta}C_{\beta\dot\beta} \,,\nn
&& \lambda^{+\alpha}\;\Rightarrow \; \theta^{+\alpha}\,l + \theta^{+\beta} l_{(\beta}^{\;\;\;\alpha)} + \bar\theta^{+\dot\beta} l_{\dot\beta}^\alpha\,, \quad
 \bar\lambda^{+\dot\alpha}\;\Rightarrow \;\bar\theta^{+\dot\alpha}\,\bar{l} + \bar\theta^{+\dot\beta} \bar l_{(\dot\beta}^{\;\;\;\dot\alpha)}
 - \theta^{+\beta}\bar l_{\beta}^{\dot\alpha}\,.
\eea
From the transformation laws \p{TranLin+} we find
\bea
&&\delta \omega_{\alpha\dot\alpha} = 2i\,l_{\alpha\dot\alpha}\,, \;\delta \bar\omega_{\alpha\dot\alpha} = -2i\,\bar l_{\alpha\dot\alpha}\,, \; \delta \omega = -2i l\,, \;
\delta \bar\omega = 2i \bar l\,, \; \delta C_{\alpha\dot\alpha} = - 2 (l_{\alpha\dot\alpha} + \bar l_{\alpha\dot\alpha}), \nn
&& \delta \Phi_{\beta\dot\beta\alpha\dot\alpha} = - 2\big(\varepsilon_{\alpha\beta} \bar{l}_{(\dot\beta\dot\alpha)} + \varepsilon_{\dot\alpha\dot\beta} l_{(\beta\alpha)}\big)\,.
\eea
Decomposing
\be
\Phi_{\beta\dot\beta\alpha\dot\alpha} = \Phi_{(\beta\alpha)(\dot\beta\dot\alpha)} + \varepsilon_{\alpha\beta} \Phi_{(\dot\beta\dot\alpha)} + \varepsilon_{\dot\alpha\dot\beta}
\Phi_{(\beta\alpha)} +  \varepsilon_{\alpha\beta}\varepsilon_{\dot\alpha\dot\beta} \Phi\,,
\ee
we observe that the complex fields $\omega_{\alpha\dot\alpha}$ and $\phi$ are purely gauge degrees of freedom: they can be put equal to zero in accordance with the WZ gauge
\p{wesszumino-grav}; using the local parameters $l_{\alpha\dot\alpha}, \bar l_{\alpha\dot\alpha}$ one can gauge away as well the spin 1 parts of the field
$\Phi_{\beta\dot\beta\alpha\dot\alpha}$, to end with the off-shell spin 2 field ($\Phi_{(\beta\alpha)(\dot\beta\dot\alpha)}, \Phi)$ and the spin 1 field $C_{\alpha\dot\alpha}$
as the only surviving physical bosonic gauge fields. The standard residual gauge transformations of these fields are associated with the local parameters $a^m(x)$ and $b(x)$ coming from
the gauge superfunctions $\lambda^m(\zeta), \lambda^5(\zeta)$:
\bea
&&    \delta_\lambda \Phi_{\beta \dot{\beta}\alpha \dot{\alpha} }
    =
    \frac{1}{2} \left(\partial_{\alpha\dot{\alpha}} a_{\beta\dot{\beta}} + \partial_{\beta\dot{\beta}} a_{\alpha\dot{\alpha}}  \right), \quad
    \delta_\lambda \Phi = \frac{1}{4} \partial_{\alpha\dot{\alpha}} a^{\alpha\dot{\alpha}}\,, \lb{spin2gauge} \\
&&\delta_\lambda C_{\alpha\dot\alpha} = -2 \partial_{\alpha\dot\alpha}\,b\,.  \lb{spin1gauge}
\eea
It was taken into account in \p{spin2gauge} that the gauge choice $\Phi_{(\beta\alpha)}=0$ (and c.c.)
expresses the parameters $l_{\alpha\dot\alpha}, \bar l_{\alpha\dot\alpha}$ as
\be
l_{\alpha\beta} = \frac{1}{4} \partial_{(\alpha\dot{\alpha}} a^{\dot{\alpha}}_{\beta)}\,, \quad l_{\dot{\alpha}\dot{\beta}}
= \frac{1}{4} \partial_{\beta(\dot{\alpha}} a^\beta_{\dot{\beta})}\,.  \lb{inducedl}
\ee

In what follows, an important role is played by the realization of the rigid ${\cal N}=2$ supersymmetry on  the superfields $h^{\pm\pm}$. Even before passing to the linearized
approximation, it is immediately seen that under the ${\cal N}=2$ transformations \p{N2SUSY} the covariantized harmonic derivatives \p{covD++}, \p{covD--} are invariant
provided the vielbeins have the following unusual transformation rules
\bea
&& \delta_\epsilon h^{++m} = -2i\big(h^{++\mu+}\sigma^m_{\mu\dot\mu} \bar\epsilon^{-\dot\mu} + \epsilon^{-\mu}\sigma^m_{\mu\dot\mu} h^{++\dot\mu+}\big)\,, \nn
&& \delta_\epsilon h^{++5} = 2i\big(h^{++\mu+}\epsilon^-_{\mu} - \bar\epsilon^{-}_{\dot\mu}h^{++\dot\mu+} \big)\,, \nn
&& \delta_\epsilon h^{++\hat\mu+} = 0\,, \lb{++susy} \\
&&\delta_\epsilon h^{--m} = -2i\big(h^{--\mu+}\sigma^m_{\mu\dot\mu} \bar\epsilon^{-\dot\mu} + \epsilon^{-\rho}\sigma^m_{\rho\dot\mu} h^{--\dot\mu+}\big)\,, \nn
&& \delta_\epsilon h^{--5} = 2i\big(h^{--\mu+}\epsilon^-_{\mu} - \bar\epsilon^{-}_{\dot\mu}h^{--\dot\mu+} \big)\,, \nn
&& \delta_\epsilon h^{--\hat\mu+} =\delta h^{--\hat\mu-} = 0\,. \lb{--susy}
\eea
These transformation laws are valid in the linearized limit too. The difference between the nonlinear and linearized cases is that in the former case
these rigid transformations form a subgroup of the gauge group \p{transf} (and its counterpart for the negatively charged vielbeins), while in the latter case they constitute an independent symmetry (which form a semi-direct product
with the relevant gauge transformations \p{TranLin-} and \p{TranLin-3}).

Now, let us define the non-analytic objects which behave as the standard $ {\cal N}=2$ superfields
\bea
&&G^{++ m} := h^{++ m} + 2i\big( h^{++\mu+}\sigma^m_{\mu\dot\mu}\bar\theta^{-\dot\mu} + \theta^{-\mu} \sigma^m_{\mu\dot\mu}h^{++\dot\mu+}\big)\,, \lb{G++} \\
&&G^{++ 5} := h^{++ 5} - 2i\big( h^{++\mu+}\theta^-_\mu - \bar\theta^-_{\dot\mu}h^{++\dot\mu+}\big)\,,\lb{G++5} \\
&&G^{-- m} := h^{-- m} + 2i\big( h^{--\mu+}\sigma^m_{\mu\dot\mu}\bar\theta^{-\dot\mu}+ \theta^{-\mu} \sigma^m_{\mu\dot\mu}h^{++\dot\mu+}\big)\,, \lb{G--} \\
&&G^{-- 5} := h^{-- 5} - 2i\big( h^{--\mu+}\theta^-_\mu - \bar\theta^-_{\dot\mu}h^{--\dot\mu+}\big)\,.\lb{G--5}
\eea
It is easy to check that\footnote{We denote by $\delta_\epsilon$ the so called passive transformations differing from the more accustomed ``active'' transformations $\delta_\epsilon^*$
by the ``transport term'', $\delta_\epsilon^* = \delta_\epsilon - \delta_\epsilon Z^M \partial_M$.}
\be
\delta_\epsilon G^{++m} = \delta_\epsilon G^{++5} = \delta_\epsilon G^{--m} = \delta_\epsilon G^{--5} = 0\,.
\ee
The newly introduced objects also possess simple transformation properties under the gauge transformations \p{TranLin+} - \p{TranLin-3}
\bea
&& \delta_\lambda G^{\pm\pm m} = D^{\pm\pm}\Lambda^m\,, \quad \delta_\lambda G^{\pm\pm 5} = D^{\pm\pm}\Lambda^5\,, \lb{TranG} \\
&& \Lambda^m = \lambda^m + 2i\big( \lambda^+\sigma^m \bar\theta^- + \theta^- \sigma^m \bar\lambda^+ \big)\,, \quad \Lambda^5 =
\lambda^5 - 2i\big( \lambda^+\theta^- - \bar\theta^- \bar\lambda^+ \big),\lb{Lambda}
\eea
and satisfy the flatness conditions
\bea
D^{++}G^{--m} = D^{--}G^{++m}\,, \quad D^{++}G^{--5} = D^{--}G^{++5} \lb{FlatG}
\eea
as a direct consequence of the harmonic equations \p{m5} - \p{spin}. The invariant linearized action of ${\cal N}=2$ SG
can be constructed just from these objects.

Let us pass to the spinor notation,
\bea
&& G^{\pm\pm \alpha\dot\alpha} =  (\tilde{\sigma}_m)^{\alpha\dot\alpha}G^{\pm\pm m}\,, \delta_{\lambda}G^{\pm\pm \alpha\dot\alpha} = D^{\pm\pm}\Lambda^{\alpha\dot\alpha}\,, \nn
&& \Lambda^{\alpha\dot\alpha} = \lambda^{\alpha\dot\alpha} + 4i\big(\lambda^{+\alpha} \bar{\theta}^{-\dot\alpha} + \theta^{-\alpha} \bar\lambda^{+\dot\alpha}\big)\,, \nn
&& G^{\pm\pm \alpha\dot\alpha} = h^{\pm\pm\alpha\dot\alpha} + 4i\big(h^{\pm\pm\alpha+} \bar{\theta}^{-\dot\alpha} + \theta^{-\alpha} \bar{h}^{\pm\pm\dot\alpha+}\big)\,, \lb{SpinNot}
\eea
and consider the manifestly ${\cal N}=2$ supersymmetric action
\be
S_1 = \int d^4x d^8\theta du \,G^{++\alpha\dot\alpha}G^{--}_{\alpha\dot\alpha}\,. \lb{S1}
\ee
Its gauge variation, with taking into account the relation \p{FlatG}, can be reduced to the expression
\be
\delta_{\lambda} S_1 = 2\int d^4xd^8\theta du\, D^{--}\Lambda^{\alpha\dot\alpha}G^{++}_{\alpha\dot\alpha}\,.\lb{1step}
\ee
Next, we pass to the integral over the analytic subspace using
\be
\int d^4xd^8\theta du = \int d\zeta^{-4} du (D^+)^4\,, \quad (D^+)^4 = \frac{1}{16} (\bar D^+)^2 (D^+)^2\,.
\ee
After some algebra, using the relation
$$
\{D^+_\alpha, \bar D^-_{\dot\alpha} \} = -2i \partial_{\alpha\dot\alpha}\,, \quad D^-_\alpha = [D^{--}, D^+_\alpha]\;\; ({\rm and \;\;c.c.})\,,
$$
as well as the property that both $\Lambda^{\alpha\dot\alpha}$ and $G^{++}_{\alpha\dot\alpha}$ are linear in $\theta^-_\alpha, \bar\theta^-_{\dot\beta}$ with analytic coefficients,
we can represent this variation as
\be
\delta_{\lambda} S_1 = 8i \int d\zeta^{-4} du \big( \partial_{\beta\dot\beta} \lambda^{+\beta}h^{++\dot\beta +} -
\partial_{\beta\dot\beta} \bar\lambda^{+\dot\beta}h^{++\beta +}\big). \lb{S11}
\ee
As the second step, we define
\be
S_2 = \int d^4xd^8\theta du \,G^{++5}G^{--5}
\ee
and, applying similar manipulations, find
\be
\delta_{\lambda} S_2= -2i \int d\zeta^{-4} du \big( \partial_{\beta\dot\beta} \lambda^{+\beta}h^{++\dot\beta +} -
\partial_{\beta\dot\beta} \bar\lambda^{+\dot\beta}h^{++\beta +}\big). \lb{S12}
\ee
So we come to the conclusion that the sum
\be
S_{(s=2)} \sim S_1 + 4 S_2 = -\frac14 \int d^4xd^8\theta du \,\big(G^{++\alpha\dot\alpha}G^{--}_{\alpha\dot\alpha} +4 G^{++5}G^{--5} \big) \lb{Spin2N2}
\ee
is invariant under both rigid ${\cal N}=2$ supersymmetry and linearized gauge transformations. So it is the invariant action of the linearized ${\cal N}=2$ SG and
the true ${\cal N}=2$ extension of the free spin 2 action. It was firstly given in \cite{Zupnik:1998td} \footnote{The linearized ${\cal N}=2$ SG in the ordinary ${\cal N}=2$ superspace
was considered in \cite{Gates:1981qq}.}. The choice of the normalization constant in this action will become clear
after considering its component bosonic sector in the WZ gauge \p{wesszumino-grav}. Note that, while proving gauge invariance of $S_1 + 4S_2$, we did not make use of
the precise structure of $G^{--}_{\alpha\dot\alpha}$ and $G^{--5}$, only the flatness conditions \p{FlatG} were employed.

\subsection{Passing to components}\lb{components}

When calculating the component action, the most annoying problem is to restore the negatively charged non-analytic gauge superfields by the basic
analytic ones $h^{++}_{\alpha\dot\beta}, h^{++5}$ and $h^{+3}_\alpha$ by using eqs.\p{m5} - \p{spin} (or \p{FlatG} for $G^{\pm\pm}_{\alpha\dot\beta}$
and $G^{\pm\pm 5}$). We shall present the full ${\cal N}=2$ component actions for any spin elsewhere; here we limit ourselves to their
bosonic sectors. Moreover, we will be basically interested in the actions for the physical gauge fields; in all cases, the auxiliary bosonic fields produce
some bilinear terms and so vanish on shell.

For the considered spin 2 case we should firstly substitute the bosonic reduction of the WZ gauge \p{wesszumino-grav}, with the additional gauge choice
$\Phi_{\alpha\dot\alpha\beta\dot\beta} = \Phi_{(\alpha\beta)(\dot\alpha\dot\beta)} + \varepsilon_{\alpha\beta}\varepsilon_{\dot\alpha\dot\beta}\Phi$,
to eqs.\p{m5} - \p{spin}. Since the latter are linear, one
can solve them separately for each term in the analytic gauge superfield. The relevant solution for the appropriate set of negatively charged gauge potentials
will contain this fixed component field together with its $x$-derivatives. As an important example, consider the following analytic monomial
\bea
h^{++ A}_{(\Phi)} = G^{++ A}_{(\Phi)} :=  i\theta^{+\beta}\theta^{+\dot\beta}\Phi_{\beta\dot\beta}^{A}\,, \lb{thetabartheta}
\eea
where the precise value of the external index $A$ is of no interest for us for the moment. For the corresponding part of the negatively charged gauge potential we obtain
\begin{multline}
    G_{(\Phi)}^{-- A}
    =
    i \theta^{-\beta} \bar{\theta}^{-\dot{\beta}} \Phi_{\beta\dot\beta}^{A}
    -
    (\theta^-)^2 \bar{\theta}^{-(\dot{\rho}} \bar{\theta}^{+\dot{\beta})}
    \partial^\beta_{\dot{\rho}} \Phi_{\beta\dot\beta}^{A}
    +
    (\bar{\theta}^-)^2 \theta^{-(\rho} \theta^{+\beta)}
    \partial_{\rho}^{\dot{\beta}}\Phi_{\beta\dot\beta}^{A}
    \\ -i(\theta^-)^2 (\bar{\theta}^-)^2 \theta^{+\rho} \bar{\theta}^{+\dot{\rho}}  \left[
    \Box \Phi_{\rho\dot\rho}^{A}
    -
    \frac12 \partial_{\rho\dot{\rho}} \partial^{\beta\dot\beta} \Phi_{\beta\dot\beta}^{A} \right]. \lb{GenForm}
\end{multline}

To find the component $(C, \;\Phi)$ action, one should also take into account that the field $P^{\alpha\dot\beta}$ in $h^{++\alpha +}$ in \p{wesszumino-grav} (and its conjugate
$\bar P^{\dot\alpha\beta}$ in $h^{++\dot\alpha +}$) is transformed under the gauge spin 2 transformations,
\be
\delta_\lambda P^{\alpha\dot\beta} = i \partial^{\dot{\beta}}_\beta l^{\alpha\beta}= -\frac{i}{4}\big( \Box a^{\alpha\dot\beta} - \frac12 \partial^{\alpha\dot\beta}\, \partial^{\gamma\dot\gamma} a_{\gamma\dot\gamma})\,,\nonumber
\ee
and so one needs to pass to the inert field $\tilde{P}_{\alpha\dot\alpha}$ through the redefinition
\begin{equation}\label{redefinition}
    P^{\mu\dot{\mu}} = \tilde{P}^{\mu\dot{\mu}} + i B^{\mu\dot{\mu}},
    \;\;\;\;\;\;\;\;\;\;
    \bar{P}^{\mu\dot{\mu}} = \tilde{\bar{P}}^{\mu\dot{\mu}} - iB^{\mu\dot{\mu}}\,,
\end{equation}
where
\begin{equation}
    B_{\beta\dot{\beta}} = \frac14\left\{3 \partial_{\beta\dot{\beta}} \Phi -  \partial^{\alpha\dot{\alpha}} \Phi_{(\alpha\beta) (\dot{\alpha}\dot{\beta})} \right\},
    \quad \delta_\lambda B_{\alpha\dot{\beta}} = -\frac14\big( \Box a_{\alpha\dot\beta} - \frac12  \partial_{\alpha\dot\beta}\, \partial^{\gamma\dot\gamma} a_{\gamma\dot\gamma}).
\end{equation}
All other auxiliary bosonic fields entering \p{wesszumino-grav} are inert under gauge transformations and all, besides the tensorial one $T^{(\mu\nu)}$, produce bilinear component
actions and so disappear on shell. The tensorial field plays an interesting role and should be retained.

Firstly we consider the part $G^{++5} G^{--5}$ in \p{Spin2N2}. The $C$-gauge field sector is determined by the analytic gauge potential
\begin{equation}
    G^{++5}_{(C)} = i \theta^{+\rho}\bar{\theta}^{+\dot{\rho}} C_{\rho\dot{\rho}}\,. \lb{++C}
\end{equation}
In accord with the general formula \p{GenForm}:
\begin{multline}
    G_{(C)}^{--5}
    =
    i \theta^{-\beta} \bar{\theta}^{-\dot{\beta}} C_{\beta\dot{\beta}}^{}
    -
    (\theta^-)^2 \bar{\theta}^{-(\dot{\rho}} \bar{\theta}^{+\dot{\beta})}
    \partial^\beta_{\dot{\rho}} C_{\beta\dot{\beta}}^{}
    +
    (\bar{\theta}^-)^2 \theta^{-(\rho} \theta^{+\beta)}
    \partial_{\rho}^{\dot{\beta}} C_{\beta\dot{\beta}}^{}
    \\ -i(\theta^-)^2 (\bar{\theta}^-)^2 \theta^{+\rho} \bar{\theta}^{+\dot{\rho}}  \left[
    \Box C^{}_{\rho\dot{\rho}}
    -
    \partial_{\rho\dot{\rho}} \partial^m C_m^{} \right].
\end{multline}

As for the tensorial auxiliary field, it  gives contributions to both $G^{++\alpha\dot\alpha}$ and $G^{++ 5}$, so we are led to compute both $G^{--\alpha\dot\alpha}$ and $G^{-- 5}$. However,
it can be shown that $G^{++\alpha\dot\alpha}_{(T)}G^{--}_{\alpha\dot\alpha(T)}$ does not contribute to the component Lagrangian, only $G^{++ 5}_{(T)}$ and $G^{-- 5}_{(T)}$ do.
For them we have the following expressions
\begin{equation}
    G_{(T)}^{++5} =
    -2i (\bar{\theta}^+)^2 \theta^+_\nu \theta^-_\mu T^{(\mu\nu)}
    -2i (\theta^+)^2 \bar{\theta}^+_{\dot{\nu}} \bar{\theta}^-_{\dot{\mu}} \bar{T}^{(\dot{\mu}\dot{\nu})}\,,
\end{equation}
\begin{multline}
    G_{(T)}^{--5}= -2i (\bar{\theta}^-)^2 \theta^+_\nu \theta^-_\mu T^{(\mu\nu)}
    -2i (\theta^-)^2 \bar{\theta}^+_{\dot{\nu}} \bar{\theta}^-_{\dot{\mu}} \bar{T}^{(\dot{\mu}\dot{\nu})}
    \\+
    2 (\bar{\theta}^-)^2 (\theta^-)^2 \bar{\theta}^{+\dot{\rho}}  \theta^+_{\mu} \partial_{\rho\dot{\rho}} T^{(\mu\rho)}
    +
    2 (\theta^-)^2 (\bar{\theta}^-)^2 \theta^{+\rho} \bar{\theta}^+_{\dot{\mu}} \partial_{\rho\dot{\nu}} \bar{T}^{(\dot{\mu}\dot{\nu})}.
\end{multline}

The total contribution of
$$G^{++5}_{(C)}G_{(C)}^{--5} + G_{(T)}^{++5}G_{(T)}^{--5} +G^{++5}_{(C)}G_{(T)}^{--5} + G_{(T)}^{++5}G_{(C)}^{--5}$$
to the component Lagrangian in \p{Spin2N2}
reads
\bea
&&\mathcal{L}_{(C,T)} = \frac14 F^{mn}F_{mn} -\big[T^{(\dot\alpha\dot\gamma)}T_{(\dot\alpha\dot\gamma)} + T^{(\alpha\gamma)} T_{(\alpha\gamma)}\big] \nn
&& \;\;\;\;\;\;\; \;\;
+\, i \big[T^{(\dot\alpha\dot\gamma)} \partial^\beta_{(\dot\alpha} C_{\beta\dot\gamma)} - T^{(\beta\rho)}\partial^{\dot\beta}_{(\beta} C_{\rho)\dot\beta}\big],\lb{Mix}
\eea
where
$$
\partial^{\dot\beta}_{(\beta} C_{\rho)\dot\beta} = \frac{i}{2} (\sigma^{mn})_{\beta\rho} F_{mn}\,, \;\partial^\beta_{(\dot\alpha} C_{\beta\dot\gamma)} =
-\frac{i}{2} (\tilde\sigma^{mn})_{\dot\alpha\dot\gamma} F_{mn}\,, \; F_{mn} = \partial_m C_n -\partial_n C_m\,.
$$
We observe the mixing between $T^{(\alpha\beta)}, T^{(\dot\alpha\dot\beta)}$ and the gauge field strength $F^{mn}$. After removing this mixing by redefining
the tensorial fields as
\bea
T_{(\alpha\beta)} = \tilde{T}_{(\alpha\beta)} + \frac{i}{2} \partial^{\dot\beta}_{(\alpha} C_{\beta)\dot\beta}, \quad T_{(\dot\alpha\dot\beta)} =\tilde{T}_{(\dot\alpha\dot\beta)}
- \frac{i}{2} \partial^{\beta}_{(\dot\alpha} C_{\beta\dot\beta)}\,
\eea
we obtain
\bea
\mathcal{L}_{(C,T)} = -\frac14 F^{mn}F_{mn} -\big[\tilde{T}^{(\dot\alpha\dot\gamma)}\tilde{T}_{(\dot\alpha\dot\gamma)} + \tilde{T}^{(\alpha\gamma)} \tilde{T}_{(\alpha\gamma)}\big].\lb{FinC}
\eea
We see that the sign of kinetic term of the gauge field has changed after this procedure and this explain the choice of the normalization factor before
the action \p{Spin2N2} \footnote{This sign is inherited from the total ${\cal N}=2$ SG action \cite{18}, where the Maxwell superfield $h^{++5}$
plays the role of compensator for the underlying gauge ${\cal N}=2$
superconformal group and, as is common for compensators,  its action has a wrong sign as compared to any other Maxwell multiplet. E.I. thanks Bernard de Wit
for useful correspondence on this issue.}.

The pure spin 2 part of the action \p{Spin2N2} is obtained from the following expressions for the pure gravitation parts of $G^{++}_{\alpha\dot\alpha}$
and $G^{++5}$ given below.

\bea
&& G_{(\Phi)}^{++\alpha\dot{\alpha}} =
    -2i \theta^{+\beta} \bar{\theta}^{+\dot{\beta}} \Phi_{\beta\dot{\beta}}^{\alpha\dot{\alpha}}
    +
    4(\theta^+)^2 \bar{\theta}^{+\dot{\beta}} \bar{\theta}^{-\dot{\alpha}} B_{\dot{\beta}}^\alpha
    -4(\bar{\theta}^+)^2 \theta^{+\beta} \theta^{-\alpha} B_\beta^{\dot{\alpha}}\,,\lb{Gphi} \\
 && G_{(\Phi)}^{++5} = -
    2 (\theta^+)^2 \bar{\theta}^{+\dot{\rho}} \theta^-_{\mu}
    B^{\mu}_{\dot{\rho}}
    -
    2 (\bar{\theta}^+)^2 \theta^{+\beta} \bar{\theta}^-_{\dot{\rho}} B^{\dot{\rho}}_{\beta}\,.\lb{G5phi}
\eea
The expressions for the relevant negatively charged potentials $G_{(\Phi)}^{--\alpha\dot{\alpha}}$ and $G_{(\Phi)}^{--5}$, are rather bulky
and are given in Appendix (eqs.\p{G--phi} and \p{G--5phi}).

After some simple though time-consuming computation we find the
contribution of $G_{(\Phi)}^{++\alpha\dot{\alpha}}G^{--}_{(\Phi)
\alpha\dot{\alpha}} + 4 G_{(\Phi)}^{++5}G_{(\Phi)}^{--5}$ to the
component spin 2 Lagrangian
\bea
&& {\cal L}_{(\Phi)} =
-\frac{1}{4}\Big[\Phi^{(\alpha\beta)(\dot{\alpha}\dot{\beta})} \Box
\Phi_{(\alpha\beta)(\dot{\alpha}\dot{\beta})} -
\Phi^{(\alpha\beta)(\dot{\alpha}\dot{\beta})}
\partial_{\alpha\dot{\alpha}} \partial^{\rho\dot{\rho}}
\Phi_{(\rho\beta)(\dot{\rho}\dot{\beta})} \nn &&
\;\;\;\;\;\;\;\;\;\;\;         + 2\, \Phi
\partial^{\alpha\dot{\alpha}} \partial^{\beta\dot{\beta}}
\Phi_{(\alpha\beta)(\dot{\alpha}\dot{\beta})}
        - 6 \Phi \Box \Phi \Big]\,.\lb{Spin2}
\eea
It is easy to check that this Lagrangian is invariant, up to a total derivative, under the gauge transformations \p{spin2gauge}. It has a correct sign agreed with that of
the spin 1 Lagrangian \p{FinC}.

So in the gauge bosonic sector we are left with the spin 2 fields $\big(\Phi^{(\alpha\beta)(\dot\alpha\dot\beta)}, \Phi \big)$ and the spin 1 field $C_{\alpha\dot\alpha}$ with the correct
Lagrangians and gauge transformations. This directly extends to the ${\cal N}=2$, ${\bf s >2}$ cases.

\section{Generalization to $\mathcal{N}=2$ spin 3 theory}

\subsection{Superfield contents and gauge symmetries}
In the ${\cal N}=2$ supersymmetric theory of the free spin 2 described above, the basic analytic superfield objects have a nice geometric meaning,
being linearized versions of the  ${\cal N}=2$ supergravity analytic supervielbein covariantizing the analyticity-preserving harmonic derivative $\mathfrak{D}^{++}$ with respect
to the superdiffeomorphism group \p{xx5} - \p{N2Diff}. For spins $s>2$  we are not aware of such a nice geometric picture. Nevertheless, it turns out that the problem of constructing
the relevant off-shell formalism can be solved just by properly generalizing the formalism of the linearized ${\cal N}=2$ supergravity described in section 2.2.

We start with ${\bf s=3}$. We introduce the real ${\cal N}=2$ bosonic superfields $h^{++(\alpha\beta)(\dot\alpha\dot\beta)} (\zeta)$,
$h^{++ \alpha\dot\alpha}(\zeta)$ (of scaling dimension $-1$) and the conjugated fermionic superfields $h^{++(\alpha\beta)\dot{\alpha}+}(\zeta)$,
$h^{++(\dot\alpha\dot\beta){\alpha}+}(\zeta)$ (of dimension $-1/2$), all being unconstrained analytic. We ascribe to them the following gauge transformation
rules as a direct generalization of \p{TranLin+}:
\bea
       && \delta h^{++(\alpha\beta)(\dot{\alpha}\dot{\beta})} = D^{++} \lambda^{(\alpha\beta)(\dot{\alpha}\dot{\beta})}
       + 4i \big[\lambda^{+(\alpha\beta)(\dot{\alpha}}  \bar{\theta}^{+\dot{\beta})}
        + \theta^{+(\alpha}  \bar{\lambda}^{+\beta)(\dot{\alpha}\dot{\beta})}\big],
        \nn
       && \delta h^{++\alpha\dot{\alpha}} = D^{++} \lambda^{\alpha\dot{\alpha}} - 2i  \big[\lambda^{+(\alpha\beta)\dot{\alpha}} \theta^{+}_{\beta} +
        \bar\lambda^{+(\dot\alpha\dot\beta){\alpha}} \bar\theta^{+}_{\dot\beta}\big], \lb{transflin2} \\
        && \delta h^{++(\alpha\beta)\dot{\alpha}+} = D^{++} \lambda^{+(\alpha\beta)\dot{\alpha}}\,,
        \nn
        &&\delta h^{++(\dot{\alpha}\dot{\beta})\alpha+} = D^{++} \bar\lambda^{+(\dot{\alpha}\dot{\beta})\alpha}\,.\label{transflin1}
\eea

Like in the ${\bf s=2}$ case, let us first to see which kind of unremovable bosonic $SU(2)$ singlet (``white'') gauge fields is retained in the newly defined gauge potentials.
 As in the case of spin 2, a simple analysis shows that all shifting $SU(2)$ singlet local symmetries are concentrated in the gauge parameters $\lambda^{+(\alpha\beta)\dot{\alpha}},
\bar\lambda^{+(\dot\alpha\dot\beta){\alpha}}$, while all bosonic gauge fields in the potentials $h^{++(\alpha\beta)(\dot{\alpha}\dot{\beta})}$ and  $h^{++\alpha\dot{\alpha}}$.
 Singling out in both sets of the objects the relevant $SU(2)$ singlet components, we find
 \bea
&&
h^{++(\alpha\beta)(\dot{\alpha}\dot{\beta})} \,\Rightarrow \, (\theta^+)^2 \omega^{(\alpha\beta)(\dot{\alpha}\dot{\beta})}
+ (\bar\theta^+)^2 \bar\omega^{(\alpha\beta)(\dot{\alpha}\dot{\beta})} -2  i\theta^+_\gamma \bar\theta^+_{\dot\gamma} \Phi^{\gamma\dot\gamma (\alpha\beta)(\dot{\alpha}\dot{\beta})}\,, \nn
&&h^{++\alpha\dot{\alpha}}\,\Rightarrow \, (\theta^+)^2\omega^{\alpha\dot{\alpha}} + (\bar\theta^+)^2\bar\omega^{\alpha\dot{\alpha}} - 2 i\theta^+_\gamma \bar\theta^+_{\dot\gamma}
C^{\gamma\dot\gamma \alpha\dot{\alpha}}\,, \nn
&& \lambda^{+(\alpha\beta)\dot{\alpha}}\,\Rightarrow \,l^{(\alpha\beta)\dot{\alpha}\gamma}\theta^+_\gamma + l^{(\alpha\beta)\dot{\alpha}\dot\gamma} \bar\theta^+_{\dot\gamma}\,, \nn
&& \bar\lambda^{+(\dot\alpha\dot\beta){\alpha}}\,\Rightarrow \,\bar l^{(\dot\alpha\dot\beta){\alpha}\dot\gamma}\bar\theta^+_{\dot\gamma} -
\bar l^{(\dot\alpha\dot\beta){\alpha}\gamma} \theta^+_{\gamma}\,.
\eea
The transformation laws \p{transflin2} imply the following gauge transformations for ``white'' component fields:
\bea
&& \delta \omega^{\alpha\dot\alpha} = i l^{(\alpha\beta) \dot\alpha}_{\;\;\;\;\;\;\;\;\;\;\beta}\,, \quad \delta \bar\omega^{\alpha\dot\alpha} =
-i \bar l^{(\dot\alpha\dot\beta)\alpha}_{\;\;\;\;\;\;\;\;\;\;\dot\beta}\,, \nn
&& \delta C^{\gamma\dot\gamma \alpha\dot\alpha} = \frac12 \big[\bar l^{(\dot\alpha\dot\gamma) \alpha\gamma} - l^{(\alpha\gamma) \dot\alpha\dot\gamma}\big]\,, \nn
&& \delta\omega^{(\alpha\beta)(\dot{\alpha}\dot{\beta})} = 2i l^{(\alpha\beta)(\dot{\alpha}\dot{\beta})}\,, \quad \delta \bar\omega^{(\alpha\beta)(\dot{\alpha}\dot{\beta})}
= -2i \bar l^{(\dot{\alpha}\dot{\beta})(\alpha\beta)}\,, \nn
&& \delta \Phi^{\gamma\dot\gamma (\alpha\beta)(\dot{\alpha}\dot{\beta})} = -2 \big[  l^{(\alpha\beta)(\dot\alpha \gamma} \varepsilon^{\dot\beta)\dot\gamma}
+ \bar l^{(\dot\alpha\dot\beta)(\alpha \dot\gamma} \varepsilon^{\beta)\gamma}\big].\nonumber
\eea
We have verified that the gauge freedom associated with the complex parameters $l^{(\alpha\beta)\dot\alpha \gamma}$ and $l^{(\alpha\beta)\dot\alpha\dot\gamma}$ (and c.c.)  is
powerful enough to gauge away fields $\omega^{\alpha\dot\alpha}, \bar\omega^{\alpha\dot\alpha}, \omega^{(\alpha\beta)(\dot\alpha\dot\beta)}, \bar\omega^{(\alpha\beta)(\dot\alpha\dot\beta)}$. Also one can use it to gauge away all the components in $C^{\gamma\dot\gamma \alpha\dot\alpha}$ apart from

\be
C^{(\gamma\alpha)(\dot\gamma\dot\alpha)} + \varepsilon^{\gamma\alpha}\varepsilon^{\dot\gamma\dot\alpha} C \lb{Ccont}
\ee
and all the components in $\Phi^{\gamma\dot\gamma(\alpha\beta)(\dot\alpha\dot\beta)}$ apart from
\be
\Phi^{(\alpha\beta\gamma)(\dot\alpha\dot\beta\dot\gamma)} + \varepsilon^{\dot\gamma(\dot\alpha} \varepsilon^{\gamma(\beta} \Phi^{\alpha)\dot\beta)}\,.\lb{Phicont}
\ee
So there survive only the pairs of fields $C^{(\gamma\alpha)(\dot\gamma\dot\alpha)}, C$ and $\Phi^{(\alpha\beta\gamma)(\dot\alpha\dot\beta\dot\gamma)},
\Phi^{\alpha\dot\beta}$ needed for the consistent description of massless spins 2 and 3, respectively  \cite{FronsdalInteg}.

Now we perform a more detailed analysis of the gauge freedom, prior to imposing any gauge on the gauge fields $C$ and $\Phi$.
This analysis leads to the following Wess-Zumino type gauge for the considered case
\bea
&&  h^{++(\alpha\beta)(\dot{\alpha}\dot{\beta})}
        =
        -2i \theta^{+\rho} \bar{\theta}^{+\dot{\rho}} \Phi^{(\alpha\beta)(\dot{\alpha}\dot{\beta})}_{\rho\dot{\rho}}
        +  (\bar{\theta}^+)^2 \theta^+ \psi^{(\alpha\beta)(\dot{\alpha}\dot{\beta})i}u^-_i  \nn
&& \;\; \;\;\;\; \;\;\;\; \;\;\;\; \;\;\; \;\;\;\; \;\;\;\;\;+\, (\theta^+)^2 \bar{\theta}^+ \bar{\psi}^{(\alpha\beta)(\dot{\alpha}\dot{\beta})i}u_i^-
        +  (\theta^+)^2 (\bar{\theta}^+)^2 V^{(\alpha\beta)(\dot{\alpha}\dot{\beta})(ij)}u^-_iu^-_j\,, \nn
&&  h^{++\alpha\dot{\alpha}} =
        -2i \theta^{+\rho} \bar{\theta}^{+\dot{\rho}} C^{\alpha\dot{\alpha}}_{\rho\dot{\rho}}
        + (\bar{\theta}^+)^2 \theta^+ \rho^{\alpha\dot{\alpha}i}u^-_i + (\theta^+)^2 \bar{\theta}^{+} \bar{\rho}^{\alpha\dot{\alpha}i}u_i^-
        + (\theta^+)^2 (\bar{\theta}^+)^2 S^{\alpha\dot{\alpha}(ij)}u^-_iu^-_j\,, \nn
&&  h^{++(\alpha\mu)\dot{\alpha}+} = (\theta^+)^2 \bar{\theta}^+_{\dot{\mu}} P^{(\alpha\mu)\dot{\alpha}\dot{\mu}}
        +  \left(\bar{\theta}^+\right)^2 \theta^+_\nu \left[\varepsilon^{\nu(\alpha} M^{\mu)\dot{\alpha}} + T^{\dot\alpha(\alpha\mu\nu)}\right]
        +  (\theta^+)^2 (\bar{\theta}^+)^2 \chi^{(\alpha\mu)\dot{\alpha}i}u^-_i\,, \nn
       &&  h^{++\alpha(\dot{\alpha}\dot{\mu})+} = \widetilde{\left(h^{++(\alpha\mu)\dot{\alpha}+}\right)}\,. \lb{WZ3}
\eea
The relevant residual gauge freedom is spanned by the following set of parameters

\begin{equation}
    \begin{cases}
        \lambda^{(\alpha\beta)(\dot{\alpha}\dot{\beta})}\; \Rightarrow \; a^{(\alpha\beta)(\dot{\alpha}\dot{\beta})}(x)\,, \\
        \lambda^{\alpha\dot{\alpha}} \;\Rightarrow\; b^{\alpha\dot{\alpha}}(x)\,,  \\
        \lambda^{(\mu\alpha)\dot{\alpha}+} \;\Rightarrow\; \epsilon^{(\mu\alpha)\dot{\alpha} i}(x) u^+_i +
        \bar{\theta}^{+\dot{\alpha}} n^{(\mu\alpha)}
        + \theta^{+\nu}l_{(\nu}^{\;\;\;\mu\alpha)\dot{\alpha}}(x)\,, \\
        \bar{\lambda}^{\alpha(\dot{\alpha}\dot{\mu})+} \;\Rightarrow\; \bar{\epsilon}^{\alpha(\dot{\alpha}\dot{\mu})  i}(x) u^+_i
        +
        \theta^{+\alpha} n^{(\dot{\alpha}\dot{\mu})}
        + \bar{\theta}^{+\dot{\nu}}l_{(\dot{\nu}}^{\;\;\;\alpha\dot{\alpha}\dot{\mu})}(x)\,. \lb{WZsymm3}
    \end{cases}
\end{equation}

These parameters are identified as:

\begin{itemize}
    \item $a^{(\alpha\beta)(\dot{\alpha}\dot{\beta})}(x)$ are local parameters of the spin 3 gauge transformations;

    \item $b^{\alpha\dot{\alpha}}(x)$ are local parameters of the spin 2 gauge transformations;

    \item $\epsilon^{(\mu\alpha)\dot{\alpha} i}(x)$ and $\bar{\epsilon}^{\alpha(\dot{\alpha}\dot{\mu})  i}(x)$ are parameters of local spin 3
    fermionic symmetry (an analog of the fermionic local symmetry for spin 2 in \p{WZsymm2});

    \item $n^{(\mu\alpha)}$ and $n^{(\dot{\alpha}\dot{\mu})}$ are parameters of local ``Lorentz rotations'' (they were present in \p{WZsymm2} as well);

    \item $l_{}^{(\nu\mu\alpha)\dot{\alpha}}(x)$ and $l_{}^{\alpha(\dot{\nu}\dot{\alpha}\dot{\mu})}(x) $ are new spin 3 analogs of the local ``Lorentz rotations''.
\end{itemize}
Note that the latter two types of parameters have been already used when coming to the irreducible contents of the bosonic gauge fields
\p{Ccont} and \p{Phicont} before attaining the complete WZ gauge. The bosonic fields $\Phi^{(\alpha\beta)(\dot{\alpha}\dot{\beta})}_{\rho\dot{\rho}},
C^{\alpha\dot{\alpha}}_{\rho\dot{\rho}}$,
and the fermionic ones $\psi^{(\alpha\beta)(\dot{\alpha}\dot{\beta})i}_\rho$ (and c.c.) are physical, the remaining fields are auxiliary. Keeping in mind the residual
gauge freedom, we are left with the full set of ${\bf 104} + {\bf 104}$  off-shell degrees of freedom.  On shell,
the multiplet $({\bf 3, 5/2, 5/2, 2})$ is retained.

The transformation laws (\ref{transflin1}) imply the following residual bosonic transformation laws:

\begin{itemize}
    \item \textbf{Spin 3 sector}

\begin{equation}
    \delta \Phi_{\beta\dot{\beta}}^{(\alpha\gamma)( \dot{\alpha}\dot{\gamma})}
    =
    \partial_{\beta\dot{\beta}} a^{(\alpha\gamma)( \dot{\alpha}\dot{\gamma})}
    -
    2 l_{(\beta}^{\;\;\;\alpha\gamma) (\dot{\alpha}} \delta^{\dot{\gamma})}_{\dot{\beta}}
    -
    2
    l^{\;\;\; \dot{\alpha}  \dot{\gamma}) (\alpha }_{(\dot{\beta}} \delta^{\gamma)}_\beta\,.
\end{equation}

We decompose the spin 3 field into the irreducible parts as:
\bea
&&    \Phi_{(\alpha\gamma) \beta (\dot{\alpha}\dot{\gamma}) \dot{\beta}}
    =
    \Phi_{(\alpha\gamma\beta ) (\dot{\alpha}\dot{\gamma}\dot{\beta} )}
    +
    \Phi_{(\alpha\gamma\beta ) (\dot{\alpha}} \varepsilon_{\dot{\gamma}) \dot{\beta} } \nn
&& \quad \quad   +\,    \Phi_{(\alpha (\dot{\alpha} \dot{\gamma} \dot{\beta} )} \varepsilon_{\gamma) \beta  }
    +
    \Phi_{(\alpha (\dot{\alpha}} \varepsilon_{\dot{\gamma}) \dot{\beta}} \varepsilon_{\gamma) \beta}\,. \lb{DecSpin3}
\eea

Using the spin 3 Lorentz transformation one can gauge away
$   \Phi_{(\alpha\gamma \beta ) \dot{\alpha} }$ and $\Phi_{\alpha (\dot{\alpha} \dot{\gamma}\dot{\beta} )}$, thus recovering the irreducible
field content \p{Phicont}:
\begin{equation}
    \Phi_{(\alpha\gamma) \beta (\dot{\alpha} \dot{\gamma}) \dot{\beta}}
    =
    \Phi_{(\alpha\gamma \beta ) (\dot{\alpha} \dot{\gamma} \dot{\beta} )}
    +
    \Phi_{(\alpha (\dot{\alpha}} \varepsilon_{\dot{\gamma}) \dot{\beta}} \varepsilon_{\gamma) \beta}\,,\lb{DecII}
\end{equation}
\begin{equation}
    \Phi_{\alpha\dot{\alpha}}
    =
    \frac{1}{9} \varepsilon^{\gamma\beta} \varepsilon^{\dot{\gamma}\dot{\beta}}
    \Phi_{(\alpha \gamma) \beta (\dot{\alpha} \dot{\gamma}) \dot{\beta}}\,. \lb{Def23}
\end{equation}

Residual spin 3 ``Lorentz'' transformations are determined from preserving the gauge $   \Phi_{(\alpha\gamma \beta ) \dot{\alpha} }
= \Phi_{\alpha (\dot{\alpha} \dot{\gamma}\dot{\beta} )} = 0$:

\bea
    \delta \Phi_{(\alpha \gamma \beta ) \dot{\alpha} } = \frac{2}{3}\partial_{(\alpha\dot{\beta}} a^{(\dot{\beta}}_{\gamma\beta) \dot{\alpha})} - 2 l_{(\alpha\gamma \beta ) \dot{\alpha} } = 0,
    \;\;\Rightarrow\;\;
    l_{(\alpha \gamma \beta ) \dot{\alpha} }
    =
    \frac{1}{3} \partial_{(\alpha\dot{\beta}} a^{(\dot{\beta}}_{\gamma\beta) \dot{\alpha})}\,,
\eea
\bea
    \delta \Phi_{\alpha (\dot{\alpha} \dot{\gamma} \dot{\beta} )}
    =
    \frac{2}{3}\partial_{\beta(\dot{\alpha}} a^{(\beta}_{\dot{\gamma}\dot{\beta}) \alpha)} - 2 l_{\alpha (\dot{\alpha} \dot{\gamma}\dot{\beta} )} = 0,
    \;\;\Rightarrow\;\;
    l_{\alpha(\dot{\alpha} \dot{\gamma} \dot{\beta} )}
    =
    \frac{1}{3}     \partial_{\beta(\dot{\alpha}} a^{(\beta}_{\dot{\gamma}\dot{\beta}) \alpha)}\,.
\eea

For the irreducible pieces in \p{DecII} we obtain the following transformations
\begin{equation}
    \delta \Phi_{(\alpha \gamma \beta ) (\dot{\alpha} \dot{\gamma} \dot{\beta} )}
    =
    \partial_{(\beta(\dot{\beta}} a_{\alpha \gamma) \dot{\alpha} \dot{\gamma})}\,, \lb{spin3gauge1}
\end{equation}
\begin{equation}
    \delta \Phi_{\alpha\dot{\beta}} = \frac{4}{9} \partial^{\gamma \dot{\gamma}} a_{(\alpha\gamma)(\dot{\beta}\dot{\gamma})}
    =
    \frac{8}{9} \partial^{m} a_{m \alpha\dot{\beta}}\,.\lb{spin3gauge2}
\end{equation}

These are the correct gauge transformation laws for the spin 3 fields.

\item \textbf{Spin 2 sector }

The transformation law of the spin 2 field $C_{\rho\dot{\rho}}^{\alpha\dot{\alpha}}$ entering the analytic potential $h^{++\alpha\dot\alpha}$ reads:

\begin{equation}
    \delta C_{\rho\dot{\rho}}^{\alpha\dot{\alpha}}= \partial_{\rho\dot{\rho}}b^{\alpha\dot{\alpha}}
    -
    n_{\dot{\rho}}^{\dot{\alpha}} \delta_\rho^\alpha + n_\rho^\alpha \delta_{\dot{\rho}}^{\dot{\alpha}}\,.
\end{equation}

After decomposing this field into the irreducible parts,

\begin{equation}
    C_{\alpha\beta \dot{\alpha} \dot{\beta}}
    =
    C_{(\alpha\beta) (\dot{\alpha} \dot{\beta})}
    +
    C_{(\alpha\beta) } \varepsilon_{\dot{\alpha} \dot{\beta}}
    +
    C_{ (\dot{\alpha} \dot{\beta})} \varepsilon_{\alpha\beta}
    +
    C \varepsilon_{\alpha\beta} \varepsilon_{\dot{\alpha} \dot{\beta}}\,,
\end{equation}

one can gauge away  $C_{(\alpha\beta) }$ and $C_{ (\dot{\alpha} \dot{\beta})}$, using local ``Lorentz'' shifts.
The residual transformations are found from preserving this ``physical'' gauge:

\begin{equation}
    2 \delta C_{(\alpha\beta)} = \partial_{(\alpha\dot{\alpha}} b^{\dot{\alpha}}_{\beta)} + 2 n_{\alpha\beta} = 0
    \;\;\;\;\;
    \Rightarrow
    \;\;\;\;\;
    n_{\alpha\beta} = - \frac{1}{2} \partial_{(\alpha\dot{\alpha}} b^{\dot{\alpha}}_{\beta)}\,, \lb{Defn3}
\end{equation}
\begin{equation}
    2 \delta C_{ (\dot{\alpha} \dot{\beta})}
    =
    \partial_{\beta(\dot{\alpha}} b^\beta_{\dot{\beta})} - 2 n_{\dot{\alpha}\dot{\beta}} = 0
    \;\;\;\;\;
    \Rightarrow
    \;\;\;\;\;
    n_{\dot{\alpha}\dot{\beta}} = \frac{1}{2} \partial_{\beta(\dot{\alpha}} b^\beta_{\dot{\beta})}\,. \lb{Defbarn3}
\end{equation}

Finally, the spin 2 field is represented as:

\begin{equation}
    C_{\alpha\beta \dot{\alpha} \dot{\beta}}
    =
    C_{(\alpha\beta) (\dot{\alpha} \dot{\beta})}
    +
    C \varepsilon_{\alpha\beta} \varepsilon_{\dot{\alpha} \dot{\beta}},
    \;\;\;\;\;\;\;\;\;\;
    C = \frac{1}{4} \epsilon^{\alpha\beta} \epsilon^{\dot{\alpha} \dot{\beta}} C_{\alpha\beta \dot{\alpha} \dot{\beta}}\,,
\end{equation}

with the following transformation laws for the constituent fields:
\begin{equation}
    \delta C_{\alpha\beta \dot{\alpha} \dot{\beta}}
    =
    \frac{1}{2} \left(\partial_{\alpha\dot{\alpha}} b_{\beta\dot{\beta}} + \partial_{\beta\dot{\beta}} b_{\alpha\dot{\alpha}}  \right),
\end{equation}
\begin{equation}
    \delta C_{(\alpha\beta) (\dot{\alpha}\dot{\beta})} = \partial_{(\beta (\dot{\beta}} b_{\alpha)\dot{\alpha})},
    \;\;\;\;\;
    \delta C = \frac{1}{4} \partial_{\alpha\dot{\alpha}} b^{\alpha\dot{\alpha}}\,.
\end{equation}
Thus the spin 2 fields have the correct transformation properties under the gauge $b_{\alpha\dot\alpha}$ symmetry.

\item \textbf{Auxiliary fields}

Like in the previous section, the bosonic auxiliary field $P^{(\alpha\mu)dot{\alpha}\dot{\mu}}$ in \p{WZ3} is not inert under
the new spin 3 Lorentz-like transformations

\begin{equation}
    \delta P^{(\alpha\mu)\dot{\alpha}\dot{\mu}} = i\partial_{\rho}^{\dot{\mu}} l_{}^{(\rho\mu\alpha)\dot{\alpha}},
    \;\;\;\;\;\;\;\;\;
    \delta \bar P^{\alpha\mu(\dot{\alpha}\dot{\mu})} =
    -i\partial^{\mu}_{\dot{\rho}} l_{}^{\alpha(\dot{\rho}\dot{\alpha}\dot{\mu})}\,.
\end{equation}

So we are led to redefine these fields to make them inert through adding proper terms depending on the spin 3 fields.
The expressions with the necessary transformation laws are as follows:
\bea
  &&  B_{(\alpha\beta) \dot{\alpha} \dot{\beta} } = -\frac{1}{2}\left\{\partial^{\gamma\dot{\gamma}} \Phi_{(\alpha \beta \gamma ) (\dot{\alpha}\dot{\gamma} \dot{\beta} )} -
    \partial_{(\alpha (\dot{\alpha}} \Phi_{\beta)\dot{\beta})}
        -
        \partial_{(\alpha\dot{\alpha}} \Phi_{\beta)\dot{\beta}}  \right\}, \nn
&& \bar{B}_{\alpha\beta (\dot{\alpha} \dot{\beta}) } = -\frac12\left\{\partial^{\gamma\dot{\gamma}} \Phi_{(\alpha \beta \gamma ) (\dot{\alpha}\dot{\gamma} \dot{\beta} )} -
    \partial_{(\alpha (\dot{\alpha}} \Phi_{\beta)\dot{\beta})}
        -
        \partial_{\alpha(\dot{\alpha}} \Phi_{\beta\dot{\beta})}  \right\}, \nn
&& \delta B_{(\alpha\beta) \dot{\alpha} \dot{\beta} }
    =
    \partial_{\rho\dot{\alpha}} l^{(\rho}_{\alpha\beta)\dot{\beta}}\,, \quad
\delta \bar{B}_{\alpha\beta (\dot{\alpha} \dot{\beta})}
    =
    \partial_{\alpha\dot{\rho}} \bar{l}^{(\dot\rho}_{\dot\alpha\dot\beta){\beta}}\,.
\eea
The sought redefinitions are:
\bea\label{transformation3}
&&P^{(\alpha\mu)\dot{\alpha}\dot{\mu}} = \tilde{P}^{(\alpha\mu)\dot{\alpha}\dot{\mu}}
+
iB^{(\alpha\mu)\dot{\alpha}\dot{\mu}},
    \quad
     \bar P^{\alpha\mu(\dot{\alpha}\dot{\mu})} = \tilde{\bar P}^{\alpha\mu(\dot{\alpha}\dot{\mu})}
     -i \bar{B}^{\alpha\mu(\dot{\alpha}\dot{\mu})}\,, \nn
&& \delta \tilde{P}^{(\alpha\mu)\dot{\alpha}\dot{\mu}} = \delta\tilde{\bar P}^{\alpha\mu(\dot{\alpha}\dot{\mu})} = 0\,.
\eea

The component fields $M^{\alpha\dot\alpha}$ and $T^{\dot\alpha(\alpha\mu\nu)}$ in \p{WZ3} have non-trivial transformation laws under the spin 2 gauge group
\bea
\delta M^{\alpha\dot\alpha} = -\frac{2}{3} i \partial_\gamma^{\dot\alpha} n^{(\alpha\gamma)}\,, \quad \delta T^{\dot\alpha(\alpha\mu\nu)} = -i \partial^{(\alpha \dot\alpha} n^{\mu\nu)}\,,
\eea
where the induced ``Lorentz'' parameters $n^{(\alpha\gamma)}$ are defined in \p{Defn3}. So these fields should also be redefined to make them inert. The redefinition required is as follows
\bea
&& T^{\dot\alpha(\alpha\mu\nu)} = \tilde{T}^{\dot\alpha(\alpha\mu\nu)} + iH^{\dot\alpha(\alpha\mu\nu)}\,, \quad M^{\alpha\dot\alpha} = \tilde{M}^{\alpha\dot\alpha}
+ iH^{\alpha\dot\alpha}\,,\nn
&& H^{\dot\alpha(\alpha\mu\nu)} = \partial_{\dot\beta}^{(\alpha}C^{\mu\nu) (\dot\alpha\dot\beta)}\,, \quad H^{\alpha\dot\alpha} = \partial^{\alpha\dot\alpha} C
- \frac13 \partial_{\beta\dot\beta}C^{(\alpha\beta)(\dot\alpha\dot\beta)}\,.\lb{redef}
\eea

\end{itemize}

\subsection{Invariant action}
To construct the invariant action for ${\cal N}=2$ spin 3 theory we need to define the negative charge non-analytic superfields analogous
to those appearing in the spin 2 case. These additional gauge potentials are
\be
h^{-- (\alpha\beta)(\dot\alpha\dot\beta)}, \quad h^{--\alpha\dot\alpha}, \quad h^{--(\alpha\beta)\dot\alpha+}, \; h^{--(\dot\alpha\dot\beta)\alpha+},
\quad h^{--(\alpha\beta)\dot\alpha-}, \quad h^{--(\dot\alpha\dot\beta)\alpha-}\,, \lb{Neg3}
\ee
and they satisfy the following harmonic equations
\bea
&& D^{++}h^{--(\alpha\beta)(\dot\alpha\dot\beta)} - D^{--}h^{++(\alpha\beta)(\dot\alpha\dot\beta)} + 4i\big[h^{--(\alpha\beta)(\dot\alpha+}\bar\theta^{+\dot\beta)}
-h^{--(\dot\alpha\dot\beta) (\alpha +}\theta^{+\beta)}\big] = 0\,, \nn
&& D^{++}h^{--\alpha\dot\beta} - D^{--}h^{++\alpha\dot\beta} -2i\big[h^{--(\alpha\beta)\dot\beta+}\theta^+_\beta -
\bar\theta^+_{\dot\alpha}h^{--(\dot\beta\dot\alpha) \alpha+}\big] = 0\,, \nn
&& D^{++}h^{--(\dot\alpha\dot\beta)\alpha+} - D^{--}h^{++(\dot\alpha\dot\beta)\alpha+} = 0\,, \;D^{++}h^{--(\alpha\beta)\dot\alpha+} - D^{--}h^{++(\alpha\beta)\dot\alpha+} = 0\,, \nn
&&D^{++}h^{--(\dot\alpha\dot\beta)\alpha-} - h^{--(\dot\alpha\dot\beta)\alpha+} =0\,, \;
D^{++}h^{--(\alpha\beta)\dot\alpha-} - h^{--(\alpha\beta)\dot\alpha+} = 0\,. \lb{HarmEq3}
\eea

These equations are covariant under the gauge transformations \p{transflin1}, provided that the negatively charged potentials are transformed as
\bea
&& \delta_\lambda h^{--(\alpha\beta)(\dot\alpha\dot\beta)} = D^{--}\lambda^{(\alpha\beta)(\dot\alpha\dot\beta)} +
4i\big[\lambda^{-(\alpha\beta)(\dot\alpha}\bar\theta^{-\dot\beta)} - \bar\lambda^{-(\dot\alpha\dot\beta) (\beta}\theta^{-\alpha)}\big]\,, \nn
&&\delta_\lambda h^{--\alpha\dot\beta} = D^{--}\lambda^{\alpha\dot\beta} - 2i  \big[\lambda^{-(\alpha\beta)\dot{\beta}} \theta^{-}_{\beta} +
        \bar\lambda^{-(\dot\alpha\dot\beta){\alpha}} \bar\theta^{-}_{\dot\alpha}\big], \nn
&& \delta_\lambda h^{--(\alpha\beta)\dot\alpha+}  = D^{--}\lambda^{+(\alpha\beta)\dot{\alpha}} - \lambda^{-(\alpha\beta)\dot{\alpha}}\,, \nn
&& \delta_\lambda h^{--(\dot\alpha\dot\beta)\alpha+}  = D^{--}\bar\lambda^{+(\dot\alpha\dot\beta){\alpha}} - \bar\lambda^{-(\dot\alpha\dot\beta){\alpha}}\,,\nn
&& \delta_\lambda h^{--(\alpha\beta)\dot\alpha-} = D^{--}\lambda^{-(\alpha\beta)\dot{\alpha}}\,, \quad \delta_\lambda h^{--(\dot\alpha\dot\beta)\alpha-} =
D^{--}\bar\lambda^{-(\dot\alpha\dot\beta){\alpha}}\,, \lb{TranNeg3}
\eea
with
\be
D^{++}\lambda^{-(\alpha\beta)\dot{\alpha}} = \lambda^{+(\alpha\beta)\dot{\alpha}}\,, \quad D^{++}\bar\lambda^{-(\dot\alpha\dot\beta){\alpha}} =
\bar \lambda^{+(\dot\alpha\dot\beta){\alpha}}\,. \lb{Condlamb3}
\ee

It is now rather straightforward to check that these harmonic equations are also covariant under the following modified rigid ${\cal N}=2$ supersymmetry
\bea
&&\delta_{\epsilon} h^{\pm\pm(\alpha\beta)(\dot\alpha\dot\beta)} = -4i\big[ h^{\pm\pm(\alpha\beta)(\dot\alpha+} \bar\epsilon^{-\dot\beta)}
- h^{\pm\pm(\dot\alpha\dot\beta) (\alpha +}\epsilon^{-\beta)}\big]\,, \nn
&&\delta_{\epsilon} h^{\pm\pm \alpha\dot\beta} = 2i \big[h^{\pm\pm(\alpha\beta)\dot\beta+}\epsilon^-_\beta - \bar\epsilon^-_{\dot\alpha}h^{\pm\pm(\dot\alpha\dot\beta)\alpha+} \big]. \lb{SUSYs3}
\eea
The passive supersymmetry variations of all other gauge potentials are vanishing, like in the spin 2 case.

The next step is to define the corresponding non-analytic objects transforming as scalar superfields under ${\cal N}=2$ supersymmetry
\bea
&& G^{\pm\pm(\alpha\beta)(\dot\alpha\dot\beta)} = h^{\pm\pm(\alpha\beta)(\dot\alpha\dot\beta)}+ 4i\big[ h^{\pm\pm(\alpha\beta)(\dot\alpha+} \bar\theta^{-\dot\beta)} -
h^{\pm\pm(\dot\alpha\dot\beta) (\alpha +}\theta^{-\beta)}\big]\,, \nn
&&G^{\pm\pm\alpha\dot\beta} = h^{\pm\pm\alpha\dot\beta} - 2i \big[h^{\pm\pm(\alpha\beta)\dot\beta+}\theta^-_\beta - \bar\theta^-_{\dot\alpha}h^{\pm\pm(\dot\alpha\dot\beta)\alpha+} \big],
\lb{DefGs3} \\
&& \delta_\epsilon G^{\pm\pm(\alpha\beta)(\dot\alpha\dot\beta)}  = \delta_\epsilon G^{\pm\pm\alpha\dot\beta} =0\,, \lb{Scalsusy3} \\
&& D^{++} G^{--(\alpha\beta)(\dot\alpha\dot\beta)} - D^{--}G^{++(\alpha\beta)(\dot\alpha\dot\beta)} = 0\,,\lb{FlatGs31} \\
&& D^{++}G^{--\alpha\dot\beta} - D^{--}G^{++\alpha\dot\beta} = 0. \lb{FlatGs3}
\eea
These superfields possess simple gauge transformation laws
\bea
&& \delta_\lambda  G^{\pm\pm(\alpha\beta)(\dot\alpha\dot\beta)} = D^{\pm\pm}\Lambda^{(\alpha\beta)(\dot\alpha\dot\beta)}\,, \quad  \delta_\lambda G^{\pm\pm\alpha\dot\beta}
= D^{\pm\pm}\Lambda^{\alpha\dot\beta}\,, \lb{GaugeG3} \\
&& \Lambda^{(\alpha\beta)(\dot\alpha\dot\beta)} = \lambda^{(\alpha\beta)(\dot\alpha\dot\beta)} + 4i\big[\lambda^{+(\alpha\beta)(\dot\alpha}\bar\theta^{-\dot\beta)}
- \bar\lambda^{+(\dot\alpha\dot\beta)(\alpha} \theta^{-\beta)} \big]\,, \lb{Lambda13} \\
&& \Lambda^{\alpha\dot\beta} = \lambda^{\alpha\dot\beta} -2i\big[\lambda^{+(\alpha\beta)\dot\beta}\theta^{-}_\beta
-\bar\theta^{-}_{\dot\alpha} \bar\lambda^{+(\dot\alpha\dot\beta)\alpha}\big]\,. \lb{Lambda23}
\eea

Passing through the same technical steps as in  sect. \ref{components}, it is a matter of direct calculation to check that
the manifestly ${\cal N}=2$ supersymmetric
action
\bea
S_{(s=3)} = \int d^4x d^8\theta du \,\Big\{G^{++(\alpha\beta)(\dot\alpha\dot\beta)}G^{--}_{(\alpha\beta)(\dot\alpha\dot\beta)}
+ 4G^{++\alpha\dot\beta}G^{--}_{\alpha\dot\beta} \Big\} \lb{Actions3}
\eea
is invariant as well under all gauge transformations and so solves the problem of finding an invariant superfield action for $ {\cal N}=2$
supersymmetric spin 3 theory. The coefficient before this invariant and its sign can be fixed by those of the spin 3 field component action.

The relevant pieces of the component action come out from the following parts of $G^{++(\alpha\beta)(\dot\alpha\dot\beta)}$ and $G^{++\alpha\dot\alpha}$
\bea
&& G_{(\Phi, B)}^{++(\alpha\beta)(\dot{\alpha}\dot{\beta})} =  -2i \theta^{+\gamma} \bar{\theta}^{+\dot{\gamma}} \Phi_{\gamma\dot{\gamma}}^{(\alpha\beta)(\dot{\alpha}\dot{\beta})}
            +
            4 (\theta^+)^2 \bar{\theta}^{+\dot{\gamma}} \bar{\theta}^{-(\dot{\alpha}} B^{\dot{\beta})(\alpha\beta)}_{\dot{\gamma}} \nn
&& -\,4 (\bar{\theta}^+)^2 \theta^{+\gamma} \theta^{-(\alpha} \bar{B}^{\beta)(\dot{\alpha}\dot{\beta})}_{\gamma}\,, \lb{G31} \\
&& G_{(\Phi, B)}^{++\alpha\dot{\alpha}} = -
        2 (\theta^+)^2 \bar{\theta}^{+\dot{\rho}} \theta^-_{\mu}
        B^{(\mu\alpha)\dot{\alpha}}_{\dot{\rho}}
        -
        2 (\bar{\theta}^+)^2 \theta^{+\beta} \bar{\theta}^-_{\dot{\rho}} \bar{B}^{\alpha(\dot{\rho}\dot{\alpha})}_{\beta}\,.\lb{G32}
\eea
Second and third terms in \p{G31} and both terms in \p{G32} follow from the redefinition \p{transformation3}.

The corresponding parts of $G_{(\Phi, B)}^{--(\alpha\beta)(\dot\alpha\dot\beta)}$ and $G_{(\Phi, B)}^{--\alpha\dot\beta}$ are given in Appendix (eqs.\p{G--31} and \p{G--32}).
Substituting all this in the superfield action \p{Actions3}, we obtain
the following component action for the spin 3 fields
\bea
&& S_{(s= 3)} =   \int d^4x \Big\{\Phi^{(\alpha_1\alpha_2\alpha_3)( \dot{\alpha}_1\dot{\alpha}_2\dot{\alpha}_3)}
 \Box \Phi_{(\alpha_1\alpha_2\alpha_3)( \dot{\alpha}_1\dot{\alpha}_2\dot{\alpha}_3)} \nn
 && \;\;\;\;\;\;\;\;\;- \,\frac{3}{2} \Phi^{(\alpha_1\alpha_2\alpha_3)( \dot{\alpha}_1\dot{\alpha}_2\dot{\alpha}_3)}
            \partial_{\alpha_1\dot{\alpha}_1} \partial^{\rho\dot{\rho}} \Phi_{(\rho\alpha_2\alpha_3)( \dot{\rho}\dot{\alpha}_2\dot{\alpha}_3)} \nn
 &&\;\;\;\;\;\;\;\;\;+ \,3 \Phi^{(\alpha_1\alpha_2\alpha_3)( \dot{\alpha}_1\dot{\alpha}_2\dot{\alpha}_3)} \partial_{\alpha_1\dot{\alpha}_1} \partial_{\alpha_2\dot{\alpha}_2}
 \Phi_{\alpha_3\dot{\alpha}_3}
            - \frac{15}{4} \Phi^{\alpha\dot{\alpha}} \Box  \Phi_{\alpha\dot{\alpha}} \nn
           && \;\;\;\;\;\;\;\;\;+\,
            \frac{3}{8} \partial_{\alpha_1\dot{\alpha}_1} \Phi^{\alpha_1\dot{\alpha}_1} \partial_{\alpha_2\dot{\alpha}_2} \Phi^{\alpha_2\dot{\alpha}_2}\Big\}\,. \lb{S3comp}
\eea
It is straightforward to check that \p{S3comp} is invariant under the spin 3 gauge group \p{spin3gauge1}, \p{spin3gauge2}. The action \p{S3comp} involves fields
$\Phi^{(\alpha_1\alpha_2\alpha_3)( \dot{\alpha}_1\dot{\alpha}_2\dot{\alpha}_3)}$ and $\Phi^{\alpha\dot{\alpha}}$ needed for the consistent description of spin 3 and coincides
with the relevant Fronsdal action. For spin 2 (which is now a superpartner of the spin 3 and is described
by the fields $C^{(\alpha\beta)(\dot\alpha\dot\beta)}, C $) also a correct Fronsdal-type action can be derived, details are given in Appendix B.

More detailed analysis of the component ${\cal N}=2$ supersymmetric spin 2 and spin 3 actions (including the fermionic contributions) will be presented elsewhere.

\section{General case: ${\cal N}=2$ integer spin s theory}
The construction described above for spins 2 and 3 can rather directly be extended to an arbitrary integer spin ${\bf s}$. Here we sketch its basic steps, without details.

The set of analytic potentials is formed by the following
analytic ${\cal N}=2$ superfields
\bea
h^{++\alpha(s-1)\dot\alpha(s-1)}(\zeta), \; h^{++\alpha(s-2)\dot\alpha(s-2)}(\zeta), \; h^{++\alpha(s-1)\dot\alpha(s-2)+}(\zeta), \;
h^{++\dot\alpha(s-1)\alpha(s-2)+}(\zeta), \lb{Sets}
\eea
where symbols $\alpha(s)$ and $\dot\alpha(s)$ denote totally symmetric combinations of $s$ spinor indices,
$\alpha(s) := (\alpha_1 \ldots \alpha_s)$,
$\dot\alpha(s) := (\dot\alpha_1 \ldots \dot\alpha_s)$. The first two potentials are bosonic, the last
two are conjugated fermionic. The corresponding gauge group
is spanned by the transformations
\bea
&& \delta_\lambda h^{++\alpha(s-1)\dot\alpha(s-1)} = D^{++} \lambda^{\alpha(s-1)\dot\alpha(s-1)} +
4i \big[\lambda^{+\alpha(s-1)(\dot\alpha(s-2)}\bar\theta^{+\dot\alpha_{s-1})} \nn
&& \;\;\;\;\;\;\;\;\;\;\;\;\;\;\;\;\;\;\;\;\;\;\;\;\;\;+\,\theta^{+(\alpha_{s-1}} \bar\lambda^{+\alpha(s-2))\dot\alpha(s-1)} \big], \nn
&&\delta_\lambda h^{++\alpha(s-2)\dot\alpha(s-2)} = D^{++} \lambda^{\alpha(s-2)\dot\alpha(s-2)} -
2i\,\big[\lambda^{+(\alpha(s-2)\alpha_{s-1})\dot\alpha(s-2)} \theta^+_{\alpha_{s-1}} \nn
&& \;\;\;\;\;\;\;\;\;\;\;\;\;\;\;\;\;\;\;\;\;\;\;\;\;\; +\,
\bar\lambda^{+(\dot\alpha(s-2)\dot\alpha_{s-1})\alpha(s-2)} \bar\theta^+_{\dot\alpha_{s-1}} \big], \nn
&& \delta_\lambda  h^{++\alpha(s-1)\dot\alpha(s-2)+} = D^{++}\lambda^{+\alpha(s-1)\dot\alpha(s-2)}\,, \nn
&& \delta_\lambda h^{++\dot\alpha(s-1)\alpha(s-2)+} =
D^{++}\bar\lambda^{+\dot\alpha(s-1)\alpha(s-2)}\,. \lb{Gauge_s}
\eea
These transformations can be used to choose the appropriate WZ gauge, like in the $s=2$ and $s=3$ cases,
and then to show that the physical multiplet involves spins
$({\bf s, s-1/2, s-1/2, s-1})$.

The next step is to define the appropriate negatively charged
potentials
\bea
h^{--\alpha(s-1)\dot\alpha(s-1)}(Z), \;
h^{--\alpha(s-2)\dot\alpha(s-2)}(Z)\,, \;
h^{--\alpha(s-1)\dot\alpha(s-2)+}(Z), \;
h^{--\dot\alpha(s-1)\alpha(s-2)+}(Z) \lb{Sets2}
\eea
(the potentials with the charges $-3$ are not essential, being fully specified by
$h^{--\alpha(s-1)\dot\alpha(s-2)+}$ and c.c.). These potentials are
related to \p{Sets} by the corresponding harmonic flatness
conditions. Then one finds that these conditions require a
non-standard realization of ${\cal N}=2$ supersymmetry on the sets
of potentials introduced. Namely,
\bea
\delta_\epsilon h^{\pm\pm\alpha(s-1)\dot\alpha(s-1)} = -4i\big[h^{\pm\pm\alpha(s-1)(\dot\alpha(s-2)+}\bar\epsilon^{-\dot\alpha_{s-1})}-
h^{\pm\pm\dot\alpha(s-1)(\alpha(s-2)+}\,\epsilon^{-\alpha_{s-1})}
\big]\,, \nn
\delta_\epsilon h^{\pm\pm\alpha(s-2)\dot\alpha(s-2)} =2i\big[h^{\pm\pm(\alpha(s-2)\alpha_{s-1})
\dot\alpha(s-2)+}\epsilon^{-}_{\alpha_{s-1}} +
h^{\pm\pm\alpha(s-2)(\dot\alpha(s-2)\dot\alpha_{s-1})+}\,\bar\epsilon^{-}_{\dot{\alpha}_{s-1}}
\big] \nonumber
\eea
(all other potentials have the standard ${\cal N}=2$ superfield ``passive'' transformation rules, e.g., $\delta_\epsilon h^{\pm\pm\alpha(s-1)\dot\alpha(s-2)+} =0$).

Next, one constructs ${\cal N}=2$ singlet
superfields
\bea
&& G^{\pm\pm\alpha(s-1)\dot\alpha(s-1)} =
h^{\pm\pm\alpha(s-1)\dot\alpha(s-1)} + 4i
\big[h^{\pm\pm\alpha(s-1)(\dot\alpha(s-2)+}\bar\theta^{-\dot\alpha_{s-1})}
\nn && \;\;\;\;\;\;\;\;\;\;\;\;\;\;\;\;\;\;\;\;\;\;\;\;\;\;-
h^{\pm\pm\dot\alpha(s-1)(\alpha(s-2)+}\,\theta^{-\alpha_{s-1})}
\big], \nn && G^{\pm\pm\alpha(s-2)\dot\alpha(s-2)} =
h^{\pm\pm\alpha(s-2)\dot\alpha(s-2)} - 2i
\big[h^{\pm\pm(\alpha(s-2)\alpha_{s-1})
\dot\alpha(s-2)+}\theta^{-}_{\alpha_{s-1}} \nn &&
\;\;\;\;\;\;\;\;\;\;\;\;\;\;\;\;\;\;\;\;\;\;\;\;\;\;+
h^{\pm\pm\alpha(s-2)(\dot\alpha(s-2)\dot\alpha_{s-1})+}\,\bar\theta^{-}_{\dot{\alpha}_{s-1}}
\big], \lb{Ggen}
\eea
which are transformed by the gauge group as
\be
\delta_\lambda G^{\pm\pm\alpha(s-1)\dot\alpha(s-1)} =
D^{\pm\pm}\Lambda^{\alpha(s-1)\dot\alpha(s-1)}\,, \quad
\delta_\lambda G^{\pm\pm\alpha(s-2)\dot\alpha(s-2)} =
D^{\pm\pm}\Lambda^{\alpha(s-2)\dot\alpha(s-2)}\,,\nonumber
\ee
where
\bea
&&\Lambda^{\alpha(s-1)\dot\alpha(s-1)} =
\lambda^{\alpha(s-1)\dot\alpha(s-1)} +
4i\big[\lambda^{+\alpha(s-1)(\dot\alpha(s-2)}\bar\theta^{-\dot\alpha_{s-1})}
- \bar\lambda^{+\dot\alpha(s-1)(\alpha(s-2)} \theta^{-\alpha_{s-1})}
\big],\nn
&&\Lambda^{\alpha(s-2)\dot\alpha(s-2)} =
\lambda^{\alpha(s-2)\dot\alpha(s-2)}
-2i\big[\lambda^{+(\alpha(s-2)\alpha_{s-1})\dot\alpha(s-2)}\theta^{-}_{\alpha_{s-1}}
\nn
&&
\;\;\;\;\;\;\;\;\;\;\;\;\;\;\;\;\;\;\;\;\;\;\;-\,\bar\theta^{-}_{\dot\alpha_{s-1}}
\bar\lambda^{+(\dot\alpha(s-2)\dot\alpha_{s-1})\alpha(s-2)}\big]\,.
\lb{LambS}
\eea

They satisfy the harmonic flatness conditions
 \bea
 && D^{++}G^{--\alpha(s-1)\dot\alpha(s-1)} = D^{--}G^{++\alpha(s-1)\dot\alpha(s-1)}\,, \nn
&& D^{++}G^{--\alpha(s-2)\dot\alpha(s-2)} = D^{--}G^{++\alpha(s-2)\dot\alpha(s-2)}\,. \nonumber
\eea

The invariant action, up to a normalization factor, is written
uniformly for any $s$:
\bea
&& S_{(s)} = (-1)^{s+1} \int d^4x
d^8\theta du \,\Big\{G^{++
\alpha(s-1)\dot\alpha(s-1)}G^{--}_{\alpha(s-1)\dot\alpha(s-1)} \nn
&&\;\;\;\;\;  +\,
4G^{++\alpha(s-2)\dot\alpha(s-2)}G^{--}_{\alpha(s-2)\dot\alpha(s-2)}
\Big\}\,. \lb{ActionsGen}
\eea
Its ${\cal N}=2$ supersymmetry is
manifest, while gauge invariance is checked by bringing the gauge
variation  to the form
\bea
&& \delta_\lambda S_{(s)} = 2(-1)^{s+1}
\int d^4x d^8\theta du
\Big\{D^{--}\Lambda^{\alpha(s-1)\dot\alpha(s-1)}G^{++
}_{\alpha(s-1)\dot\alpha(s-1)} \nn &&\;\;\;\;\;\;\;  +\, 4
D^{--}\Lambda^{\alpha(s-2)\dot\alpha(s-2)}G^{++
}_{\alpha(s-2)\dot\alpha(s-2)}\Big\}
\eea
and further proceeding as
in the check of invariance of the actions \p{Spin2N2} and
\p{Actions3}.  Finally, one gets $\delta_\lambda S_{(s)}=0$.

The component actions can be deduced from \p{ActionsGen} by means of the
same tools as those used when deriving the component actions for the
spin 3 case. In the WZ gauge the basic bosonic gauge fields are
contained in the analytic potentials
$h^{++\alpha(s-1)\dot\alpha(s-1)}$ and
$h^{++\alpha(s-2)\dot\alpha(s-2)}$,
\bea
&& h^{++\alpha(s-1)\dot\alpha(s-1)} = -2i
\theta^{+\alpha_s}\bar\theta^{+\dot\alpha_s}
\Phi^{\alpha(s-1)\dot\alpha(s-1)}_{\alpha_s \dot\alpha_s} + \ldots,
\nn && h^{++\alpha(s-2)\dot\alpha(s-2)}= -2i
\theta^{+\alpha_{s-1}}\bar\theta^{+\dot\alpha_{s-1}}
C^{\alpha(s-2)\dot\alpha(s-2)}_{\alpha_{s-1} \dot\alpha_{s-1}} +
\ldots\,. \lb{Spins}
\eea
The residual gauge freedom in the WZ
gauge proves to be so powerful that it allows one to remove from the
gauge fields $\Phi^{\alpha(s-1)\dot\alpha(s-1)}_{\alpha_s
\dot\alpha_s}$ and $C^{\alpha(s-2)\dot\alpha(s-2)}_{\alpha_{s-1}
\dot\alpha_{s-1}}$ all the irreducible components except for \bea
\big\{\Phi^{\alpha(s)\dot\alpha(s)},  \quad
\Phi^{\alpha(s-2)\dot\alpha(s-2)} \big\},
\quad\big\{C^{\alpha(s-1)\dot\alpha(s-1)},  \quad
C^{\alpha(s-3)\dot\alpha(s-3)} \big\}\,,
\lb{Spin-s-fields}
\eea
which are just pairs
of tensor fields needed for the consistent off-shell description of
the massless spins ${\bf s}$ and ${\bf s-1}$ in the Fronsdal approach \footnote{Since the on-shell ${\cal N}=2$
supermultiplet contains two integer spins, ${\bf s}$ and ${\bf s-1}$, after reductions to components we naturally
obtain  a sum of Fronsdal actions for spins ${\bf s}$ and ${\bf s-1}$. }. Their
correct gauge transformation laws can easily be derived from the
superfield ones on the pattern of the previously considered ${\cal
N}=2$ supersymmetric spin ${\bf  s=2}$ and spin ${\bf s=3}$ models.

The gauge freedom allowing to gauge away all the ``white'' ($SU(2)$ singlet) bosonic components from the basic gauge super potentials
beyond those in \p{Spin-s-fields} is contained in the following  pieces of the spinor gauge superfunctions
 $\lambda^{+\alpha(s-1)\dot\alpha(s-2)}, \bar\lambda^{+\alpha(s-2)\dot\alpha(s-1)}$:
\bea
\lambda^{+\alpha(s-1)\dot\alpha(s-2)} \;\Rightarrow \; \omega^{\alpha(s-1)\beta\dot\alpha(s-2)}\theta^+_\beta
+ \omega^{\alpha(s-1)\dot\alpha(s-2)\dot\beta}\bar\theta^+_{\dot\beta}\,, \;\;(\rm and\;\, c.c.)\,.
\eea

\section{Summary and outlook}
In this paper we presented an off-shell ${\cal N}=2$ supersymmetric
extension of the Fronsdal theory \cite{FronsdalInteg} for integer
spins in terms of unconstrained ${\cal N}=2$ superfields. For any spin ${\bf s}\geq 2$ the relevant off-shell multiplet is described by
a triad of unconstrained harmonic analytic superfields
$h^{++\alpha(s-1)\dot\alpha(s-1)}(\zeta)$,
$h^{++\alpha(s-2)\dot\alpha(s-2)}(\zeta)$ and
$h^{++\alpha(s-1)\dot\alpha(s-2)+}(\zeta)$ (and c.c.), which are
subjected to gauge transformations with the analytic superfield
parameters. The on-shell content of the spin ${\bf s}$ multiplet is $({\bf
s, s-1/2, s-1/2, s-1})$\footnote{One can include the spin ${\bf s =1}$
into this hierarchy as well:  it is described by a single analytic
superfield $h^{++ 5}$ and encompasses the Abelian gauge ${\cal N}=2$
multiplet (spins $({\bf 1, 1/2, 1/2, 0})$ on shell). Note that the off-shell contents of ${\cal N}=2$ multiplets with ${\bf s=1, 2, 3}$ as the higher
spins amount to $n_{\bf s} = 2 \, \times\, 8\big[ {\bf s}^2 + ({\bf s-1})^2\big]$ essential degrees of freedom. It would be interesting to derive this universal formula for any
spin ${\bf s}$ from a purely group-theoretical consideration.}.  For these
superfields we  found the ${\cal N}=2$ supersymmetric and gauge
invariant superfield actions which surprisingly have the universal
form \p{ActionsGen}. For ${\bf s =2}$ and ${\bf s=3}$ we have explicitly shown
that this off-shell superfield action yields the correct gauge
invariant actions for the spin ${\bf s}$ and ${\bf s-1}$ components of the relevant
multiplets.

These findings raise a lot of problems we are going to address in the nearest future.

\begin{itemize}
\item  A natural next step would be construction of an analogous ${\cal N}=2$ supersymmetric extensions
of theories with the half-integer highest spin
\cite{FronsdalHalfint};

\item We would like also to learn how the harmonic superspace construction  could be extended to
AdS (and more general conformally-flat) space-time backgrounds;

\item  It is of interest to explore possible relationships with the ${\cal N}=2$ superconformal
higher spins which recently received some attention
\cite{Kuzenko:2017ujh}, \cite{Buchbinder:2019yhl}, \cite{KPR},
\cite{Kuzenko:2021pqm}. ${\cal N}=2$ conformal supergravity also
admits a geometric formulation in HSS \cite{Galperin:1987ek,18}, so
it is natural to expect that there exist some higher-spin HSS models
generalizing the linearized version of such a formulation;

\item  There exist a few non-equivalent off-shell versions of Einstein ${\cal N}=2$ SG related to
different choices of the superconformal compensator for ${\cal
N}=2$ Weyl multiplet. Our formulation of ${\cal N}=2$ higher spins
is built on a generalization of the minimal version.  It would be
interesting to construct analogous off-shell formulations (if
exist), proceeding from the linearizations of  other versions of Einstein
${\cal N}=2$ SG;

\item As usual, the most difficult problem would be constructing a self-consistent interacting theory with the free actions presented here as a point of departure, and finding out
 appropriate deformations of the higher-spin superfield gauge symmetries. As a first step towards this goal one could attempt to couple the theory to full ${\cal N}=2$ Einstein
 supergravity by replacing the flat harmonic derivatives $D^{\pm\pm}$ altogether by the covariantized ones $\mathfrak{D}^{\pm\pm}$, though for the time being it is
 unclear how to generalize the action \p{ActionsGen}. Anyway, the interactions will involve the same off-shell analytic harmonic superfields as the free theory discussed here.
 It is highly likely that the interaction case will require considering at once an infinite sequence of such actions (with all spins),
 in accord with the well-known Fradkin-Vasiliev arguments \cite{FradVas,Vasiliev:1990en};

 \item A related problem is to couple the higher ${\cal N}=2$ spins to the hypermultiplet matter to
 which all other matter ${\cal N}=2$ multiplets are related by the proper
 superfield duality transformations \cite{18};

 \item It is known that the $4D, {\cal N}=4$ super Yang-Mills theory can be
 formulated in terms of $4D, {\cal N}=2$ harmonic superfields as a
 theory of coupled ${\cal N}=2$ vector multiplet and hypermultiplet \cite{18}.
 Based on this analogy, one can hope that it will be possible to construct ${\cal N}=4$
 supersymmetric higher-spin theory in terms of proper ${\cal N}=2$
 harmonic superfields;

\item At last, it is interesting to work out an analogous harmonic superspace setting for higher spins with extended supersymmetry
in other dimensions (e.g., in $6D$ case).

\end{itemize}

The above formulation suggests the following geometric conjecture. As was already pointed out, the basic analytic potentials of the ${\bf s=2}$ case originate from the analytic
vielbein of ${\cal N}=2$ supergravity in HSS which covariantizes the analyticity-preserving harmonic derivative $\mathfrak{D}^{++}$ and their index structure matches with that of  the derivatives
$\frac{\partial}{\partial x^{\mu\dot\mu}}$ and $\frac{\partial}{\partial \theta^{\mu, \dot\mu}}$ inside $\mathfrak{D}^{++}$. Then it is natural to assume that
the higher-spin analogs of these potentials could be associated with some non-trivial extensions of the standard superspace by new tensorial and spinorial coordinates of the type
$x^{(\alpha\beta)(\dot\alpha\dot\beta)}, \theta^{+ (\alpha\beta)\dot\alpha}, \bar\theta^{+ (\dot\alpha\dot\beta)\alpha}$ (and their multi-index analogs). In the complete hypothetical
supergravity-type theory, the gauge functions like $\lambda^{(\alpha\beta)(\dot\alpha\dot\beta)}\,,  \lambda^{(\alpha\beta)\dot\alpha}$ could geometrically  appear as local shifts
of these new coordinates.
Also, the plenty of spinor indices $\alpha, \beta, \dot\alpha, \dot\beta \ldots$ which characterize the basic objects of the theories considered could seemingly be hidden
by introducing the commuting twistor-like spinorial variables $\tau_\alpha, \bar\tau_{\dot\alpha}$ and contracting the spinor indices
with them. Adding such extra variables could essentially facilitate  dealing with various objects of the ${\cal N}=2$ higher-spin
theories constructed and their various generalizations, even though the geometric  meaning of such variables within the present context is as yet unclear.


\section*{Acknowledgements}
We thank M. Vasiliev for useful discussion and comments and S.~Kuzenko for bringing to our attention ref. \cite{SeSi}. Work of I.~B. and E.~I.
was supported in part by the Ministry of Education of Russian Federation,
project FEWF-2020-0003.


\vspace{1.5cm}

\renewcommand\theequation{A.\arabic{equation}} \setcounter{equation}0
\subsection*{A\quad Some technical issues}
\vspace{0.4cm}

\noindent{\bf Spin 2 sector}\\

Here we present the negatively charged potentials for $G^{++\alpha\dot\alpha}_{(\Phi)}$ and  $G^{++5}_{(\Phi)}$ defined in eqs. \p{Gphi} and \p{G5phi}.
The expressions for them are as follows
\bea
&&G_{(\Phi)}^{--\alpha\dot{\alpha}}
    =
    -2i \theta^{-\beta} \bar{\theta}^{-\dot{\beta}} \Phi_{\beta\dot{\beta}}^{\alpha\dot{\alpha}}
    +
    2 (\theta^-)^2 \bar{\theta}^{-(\dot{\rho}} \bar{\theta}^{+\dot{\beta})}
    \partial^\beta_{\dot{\rho}} \Phi_{\beta\dot{\beta}}^{\alpha\dot{\alpha}}
    -
    2 (\bar{\theta}^-)^2 \theta^{-(\rho} \theta^{+\beta)}
    \partial_{\rho}^{\dot{\beta}} \Phi_{\beta\dot{\beta}}^{\alpha\dot{\alpha}} \nn
&& \;\;\;\;\;\;\;\;\; +\, 4(\theta^-)^2 \bar{\theta}^{+\dot{\alpha}} \bar{\theta}^{-\dot{\beta}} B_{\dot{\beta}}^{\alpha}
    -
    4 (\bar{\theta}^-)^2 \theta^{+\beta} \theta^{-\alpha} B^{\dot{\alpha}}_{\beta}
    - 4i(\theta^-)^2 (\bar{\theta}^-)^2 \theta^{+\rho} \bar{\theta}^{+\dot{\rho}}  \mathcal{G}_{\rho\dot{\rho}}^{\alpha\dot{\alpha}}\,,\lb{G--phi}\\
&& G_{(\Phi)}^{--5} =
    2 (\theta^-)^2 \bar{\theta}^{-\dot{\rho}} \theta^+_{\mu}
    B^{\mu}_{\dot{\rho}}
    +
    i (\theta^+)^2 (\bar{\theta}^-)^2 (\theta^-)^2  \partial_{\rho\dot{\rho}} B^{\rho\dot{\rho}} \nn
&& \;\;\;\;\;\;\;\;\;   + 2\, (\bar{\theta}^-)^2 \theta^{-\beta} \bar{\theta}^+_{\dot{\rho}} B_\beta^{\dot{\rho}}
    -
    i (\bar{\theta}^+)^2 (\bar{\theta}^-)^2 (\theta^-)^2 \partial_{\rho\dot{\rho}} B^{\rho\dot{\rho}}\,.\lb{G--5phi}
\eea

In eq. \p{G--phi} the following notation was used
\begin{equation}\label{lin-ein}
    \mathcal{G}_{\alpha\beta\dot{\alpha}\dot{\beta}}
    =
    \mathcal{R}_{(\alpha\beta)(\dot{\alpha}\dot{\beta})} - \varepsilon_{\alpha\beta} \varepsilon_{\dot{\alpha}\dot{\beta}} \mathcal{R}
\end{equation}
and
\bea
&& \mathcal{R}_{(\alpha\beta)(\dot{\alpha}\dot{\beta})} =
        \frac{1}{2} \partial_{(\alpha(\dot{\alpha}} \partial^{\sigma\dot{\sigma}} \Phi_{\beta)\sigma\dot{\beta})\dot{\sigma}}
        -
        \frac{1}{2} \Box \Phi_{(\alpha\beta)(\dot{\alpha}\dot{\beta})}
        -
        \frac{1}{2} \partial_{(\alpha(\dot{\alpha}} \partial_{\beta)\dot{\beta})} \Phi \, ,\nn
&& \mathcal{R} = \frac{1}{8} \partial^{\alpha\dot{\alpha}} \partial^{\beta\dot{\beta}}  \Phi_{(\alpha\beta) (\dot{\alpha}\dot{\beta})}
        - \frac{3}{4} \Box \Phi \,. \nonumber
\eea

The object $\mathcal{G}_{\alpha\beta\dot{\alpha}\dot{\beta}}$ is just the linearized form of Einstein tensor,
$\mathcal{R}_{(\alpha\beta)(\dot{\alpha}\dot{\beta})}$ and $\mathcal{R}$ are related to the linearized scalar curvature $R$ and Ricci tensor $R_{mn}$ in the tensor notation as:
\bea
&& \mathcal{R}_{(\alpha\beta)(\dot{\alpha}\dot{\beta})} +
    \mathcal{R}\; \epsilon_{\alpha\beta} \epsilon_{\dot{\alpha}\dot{\beta}} = (\sigma^m)_{\alpha\dot{\alpha}} (\sigma^n)_{\beta\dot{\beta}} R_{(mn)}\,, \quad \mathcal{R} = \frac{1}{2} R\,, \nn
&& R = \partial^m \partial^n h_{mn} - \Box h\,, \quad R_{mn} = \frac{1}{2} (\partial^k \partial_m h_{nk} + \partial^k \partial_n h_{mk} - \Box h_{mn} - \partial_m \partial_n h), \nonumber
\eea
where
\be
 h_{mn} = \frac14 (\tilde{\sigma}_m)^{\dot\alpha \alpha}(\tilde{\sigma}_n)^{\dot\beta \beta}\big[\Phi_{(\alpha\beta)(\dot{\alpha}\dot{\beta})}
        +
        \Phi \varepsilon_{\alpha\beta} \varepsilon_{\dot{\alpha}\dot{\beta}}]. \nonumber
\ee
It is straightforward to check that the expressions \p{G--phi} and \p{G--5phi}, together with \p{Gphi} and \p{G5phi}, solve the harmonic flatness conditions \p{FlatG}.

\vspace{0.5cm}

\noindent{\bf Spin 3 sector}\\

Here we present the negatively charged potentials for $G^{++(\alpha\beta)(\dot{\alpha}\dot{\beta})}_{(\Phi, B)}$ and  $G^{++\alpha\dot{\alpha}}_{(\Phi,B)}$ defined in eqs. \p{G31} and \p{G32}.
The expressions for them are as follows
\bea
&&G_{(\Phi, B)}^{--(\alpha\beta)(\dot{\alpha}\dot{\beta})}
=
-2i \theta^{-\beta} \bar{\theta}^{-\dot{\beta}} \Phi_{\beta\dot{\beta}}^{(\alpha\beta)(\dot{\alpha}\dot{\beta})}
\nn
&& \;\;\;\;\;\;\;\;\;
+
2 (\theta^-)^2 \bar{\theta}^{-(\dot{\rho}} \bar{\theta}^{+\dot{\beta})}
\partial^\beta_{\dot{\rho}} \Phi_{\beta\dot{\beta}}^{(\alpha\beta)(\dot{\alpha}\dot{\beta})}
-
2 (\bar{\theta}^-)^2 \theta^{-(\rho} \theta^{+\beta)}
\partial_{\rho}^{\dot{\beta}} \Phi_{\beta\dot{\beta}}^{(\alpha\beta)(\dot{\alpha}\dot{\beta})} \nn
&& \;\;\;\;\;\;\;\;\; +\, 4(\theta^-)^2 \bar{\theta}^{+(\dot{\alpha}} \bar{\theta}^{-\dot{\rho}} B_{\dot{\rho}}^{\dot{\beta})(\alpha\beta)}
-
4 (\bar{\theta}^-)^2 \theta^{+\rho} \theta^{-(\alpha} \bar{B}^{\beta)(\dot{\alpha}\dot{\beta})}_{\rho}
\nn
&& \;\;\;\;\;\;\;\;\;
- 3i(\theta^-)^2 (\bar{\theta}^-)^2 \theta^{+\rho} \bar{\theta}^{+\dot{\rho}}  \mathcal{G}_{\rho\dot{\rho}}^{(\alpha\beta)(\dot{\alpha}\dot{\beta})}\,,\lb{G--31}\\
&& G_{(\Phi, B)}^{--\alpha\dot{\alpha}} =
2 (\theta^-)^2 \bar{\theta}^{-\dot{\rho}} \theta^+_{\mu}
B^{(\mu\alpha)\dot{\alpha}}_{\dot{\rho}}
+
i (\theta^+)^2 (\bar{\theta}^-)^2 (\theta^-)^2  \partial_{\rho\dot{\rho}} B^{(\rho\alpha)\dot{\rho}\dot{\alpha}} \nn
&& \;\;\;\;\;\;\;\;\;   + 2\, (\bar{\theta}^-)^2 \theta^{-\beta} \bar{\theta}^+_{\dot{\rho}} \bar{B}_\beta^{\alpha(\dot{\alpha}\dot{\rho})}
-
i (\bar{\theta}^+)^2 (\bar{\theta}^-)^2 (\theta^-)^2 \partial_{\rho\dot{\rho}} \bar{B}^{\rho\alpha(\dot{\rho}\dot{\alpha})}\,.\lb{G--32}
\eea

In eq. \p{G--31} the following notation was used:
\bea\label{Eintein3}
\mathcal{G}_{(\alpha_1\alpha_2)\alpha_3 (\dot{\alpha}_1  \dot{\alpha}_2) \dot{\alpha}_3}
=
\mathcal{R}_{(\alpha_1\alpha_2\alpha_3) (\dot{\alpha}_1  \dot{\alpha}_2 \dot{\alpha}_3)}
- \frac{4}{9}
\mathcal{R}_{(\alpha_1 (\dot{\alpha}_1} \epsilon_{\dot{\alpha}_2) \dot{\beta}} \epsilon_{\alpha_2) \beta}
\eea

and
\begin{equation}
    \mathcal{R}_{\alpha_1\dot{\alpha}_1}
    =
    \partial^{\alpha_2\dot{\alpha}_2} \partial^{\alpha_3\dot{\alpha}_3} \Phi_{(\alpha_1\alpha_2\alpha_3)( \dot{\alpha}_1\dot{\alpha}_2\dot{\alpha}_3)}
    -
    \frac{1}{4}\partial_{\alpha_1\dot{\alpha}_1} \partial^{\alpha_2\dot{\alpha}_2} \Phi_{\alpha_2\dot{\alpha}_2}
    -
    \frac{5}{2} \Box \Phi_{\alpha_1\dot{\alpha}_1}\,,
\end{equation}
\begin{eqnarray}
    \mathcal{R}_{(\alpha_1\alpha_2\alpha_3)( \dot{\alpha}_1\dot{\alpha}_2\dot{\alpha}_3)}
    &=&
    \partial_{(\alpha_1(\dot{\alpha}_1}\partial^{\rho\dot{\rho}} \Phi_{\alpha_2\alpha_3)\rho \dot{\alpha}_2\dot{\alpha}_3)\dot{\rho}}
    -
    \frac{2}{3}
    \Box \Phi_{(\alpha_1\alpha_2\alpha_3)( \dot{\alpha}_1\dot{\alpha}_2\dot{\alpha}_3)} \nn
    &&-\,
    \partial_{(\alpha_1(\dot{\alpha}_1}
    \partial_{\alpha_2\dot{\alpha}_2}
    \Phi_{\alpha_3)\dot{\alpha}_3)}\,.
\end{eqnarray}

\subsection*{B\quad Spin 2 sector of $\mathcal{N}=2$ spin 3 theory}

Here we present the relevant pieces of the analytic  gauge potential
in the spin 2 sector of  $\mathcal{N}=2$ spin 3 theory and the corresponding parts of the negatively
charged  potentials. Using them,  we derive the component form  of
the  superfield action \p{Actions3} in the spin 2 sector.

As a consequence  of the redefinition of \p{redef} and relations \p{DefGs3} we have in the spin 2 sector:

\begin{equation}
    G_{(s=2)}^{++(\alpha\beta)(\dot{\alpha}\dot{\beta})}
    =
    - 4 (\bar{\theta}^+)^2 \theta^+_\rho \bar{\theta}^{-(\dot{\alpha}} H^{\dot{\beta}) \rho(\alpha\beta)}
    +
    4 (\theta^+)^2 \bar{\theta}^+_{\dot{\rho}} \theta^{-(\alpha}
    \bar{H}^{\beta)\dot{\rho}(\dot{\alpha}\dot{\beta})}\,,
\end{equation}
\begin{multline}
    G^{++\alpha\dot{\alpha}}_{(s=2)}
    =
    -2 i \theta^{+\rho} \bar{\theta}^{+\dot{\rho}} C_{\rho\dot{\rho}}^{\alpha\dot{\alpha}}
    +
    2 (\theta^+)^2 \bar{\theta}^{+}_{\dot{\rho}} \bar{\theta}^{-}_{\dot{\mu}} \bar{H}_{}^{\alpha\dot{\rho}(\dot{\mu}\dot{\alpha})}
    +
    2 (\bar{\theta}^+)^2 \theta^{+}_{\rho}\theta^{-}_{\mu} H_{}^{\dot{\alpha}\rho(\alpha\mu)}
    \,.
\end{multline}

The negatively charged potentials can be obtained as a solution of eqs. \p{FlatGs31} and \p{FlatGs3}:

\begin{equation}
    G_{(s=2)}^{--(\alpha\beta)(\dot{\alpha}\dot{\beta})}
    =
    - 4 (\bar{\theta}^-)^2 \theta^-_\rho \bar{\theta}^{+(\dot{\alpha}} H^{\dot{\beta}) \rho(\alpha\beta)}
    +
    4 (\theta^-)^2 \bar{\theta}^-_{\dot{\rho}} \theta^{+(\alpha}
    \bar{H}^{\beta)\dot{\rho}(\dot{\alpha}\dot{\beta})}
    +
    \dots\,,
\end{equation}
\begin{multline}
    G_{(s=2)}^{--\alpha\dot{\alpha}} = -2 i \theta^{-\rho} \bar{\theta}^{-\dot{\rho}} C_{\rho\dot{\rho}}^{\alpha\dot{\alpha}}
    +2
    (\theta^-)^2 \bar{\theta}^{-(\dot{\rho}} \bar{\theta}^{+\dot{\beta})}
    \partial^\beta_{\dot{\rho}} C_{\beta\dot{\beta}}^{\alpha\dot{\alpha}}
    -2
    (\bar{\theta}^-)^2 \theta^{-(\rho} \theta^{+\beta)}
    \partial_{\rho}^{\dot{\beta}} C_{\beta\dot{\beta}}^{\alpha\dot{\alpha}}
    \\
    +
    2 (\theta^-)^2 \bar{\theta}^{+}_{\dot{\rho}} \bar{\theta}^{-}_{\dot{\mu}} \bar{H}_{}^{\alpha\dot{\rho}(\dot{\mu}\dot{\alpha})}
    +2 (\bar{\theta}^-)^2 \theta^{+}_{\rho} \theta^{-}_{\mu} H_{}^{\dot{\alpha}\rho(\mu\alpha)}
    +4
    i(\theta^-)^2 (\bar{\theta}^-)^2 \theta^{+}_{\rho} \bar{\theta}^{+}_{\dot{\rho}}
    \mathcal{G}^{\alpha\rho\dot{\alpha}\dot{\rho}}\,.
\end{multline}

Here, $\mathcal{G}^{\alpha\rho\dot{\alpha}\dot{\rho}}$ is the linearized form of Einstein tensor \p{lin-ein}. We also used the notations:
\begin{equation*}
    \bar{H}^{\alpha\dot{\rho}(\dot{\alpha}\dot{\mu})}
    :=
    \bar{H}^{\alpha(\dot{\rho}\dot{\alpha}\dot{\mu})}
    +
    \epsilon^{\dot{\rho}(\dot{\alpha}} \bar{H}^{\alpha\dot{\mu})}\,,
    \;\;\;\;\;\;\;\;\;
    H^{\dot{\alpha}\rho(\mu\alpha)}
    :=
    H^{\dot{\alpha}(\rho\mu\alpha)}
    +
    \epsilon^{\rho(\alpha} H^{\dot{\alpha}\mu) }\,,
\end{equation*}
\begin{equation*}
    \bar{H}^{\alpha(\dot{\rho}\dot{\alpha}\dot{\mu})}
        =
        -\partial_\beta^{(\dot{\rho}}C^{\dot{\alpha}\dot{\mu})(\alpha\beta)}\,,
    \;\;\;\;\;\;\;\;
    H^{\dot{\alpha}(\rho\mu\alpha)}
        =
        \partial^{(\alpha}_{\dot{\beta}} C^{\mu\rho)(\dot{\alpha}\dot{\beta})}\,,
\end{equation*}
\begin{equation*}
    H^{\mu\dot{\mu}}
        =
        \partial^{\mu\dot{\mu}} C
        -
        \frac{1}{3} \partial_{\rho\dot{\rho}} C^{(\mu\rho)(\dot{\mu}\dot{\rho})}
        =
        -   \bar{H}^{\mu\dot{\mu}}\,.
\end{equation*}

Substituting all this in the superfield action \p{Actions3}, we
obtain the spin 2 action of the $\mathcal{N}=2$ spin 3 theory

\begin{multline}
    S_{(s=2)}
    =
    16\int d^4x \; C^{\alpha\beta\dot{\alpha}\dot{\beta}}
    \mathcal{G}_{\alpha\beta\dot{\alpha}\dot{\beta}}
    =
    -8 \int d^4x \; \Big[C^{(\alpha\beta)(\dot{\alpha}\dot{\beta})} \Box
    C_{(\alpha\beta)(\dot{\alpha}\dot{\beta})}
    \\-
    C^{(\alpha\beta)(\dot{\alpha}\dot{\beta})}
    \partial_{\alpha\dot{\alpha}} \partial^{\rho\dot{\rho}}
    C_{(\rho\beta)(\dot{\rho}\dot{\beta})}  + 2\, C
    \partial^{\alpha\dot{\alpha}} \partial^{\beta\dot{\beta}}
    C_{(\alpha\beta)(\dot{\alpha}\dot{\beta})}
    - 6 C \Box C \Big].
\end{multline}

This action is the linearized Einstein action and, up to a normalization factor, coincides with
the action  corresponding to the Lagrangian \p{Spin2}.



\begin{thebibliography}{99}
\bibitem{GGRS} S.~J.~Gates, Jr, M.~T.~Grisaru, M.~Ro{\u c}ek, W.~Siegel, {\it Superspace, or One Thousand and
One Lessons in Supersymmetry}, Benjamin Cummings, Reading, MA, 1983,
548 p.,  arXiv:hep-th/0108200.

\bibitem{BL} I.~L.~Buchbinder, S.~M.~Kuzenko, {\it Ideas and Methods of
Supersymmetry and Supergravity or a Walk Through Superspace}, IOP
Publishing, 1998, 656 p.

\bibitem{18} A.~S.~Galperin, E.~A.~Ivanov, V.~I.~Ogievetsky, E.~S.~Sokatchev,
{\it Harmonic superspace}, Cambridge Monographs on Mathematical
Physics, Cambridge University Press, 2001, 306 p.



\bibitem{HSS} A.~Galperin, E.~Ivanov, V.~Ogievetsky, E.~Sokatchev, {\it Harmonic superspace: key to $N=2$ supersymmetric theories},
Pis'ma ZhETF {\bf 40} (1984) 155 [JETP Lett. {\bf 40} (1984) 912];
A.~S.~Galperin, E.~A.~Ivanov, S.~Kalitzin, V.~I.~Ogievetsky, E.~S.~Sokatchev,{\it Unconstrained ${\cal N}=2$ Matter, Yang-Mills
and Supergravity Theories in Harmonic Superspace}, Class. Quant. Grav. {\bf 1} (1984) 469-498 [Erratum: Class. Quant. Grav. {\bf 2}
(1985) 127].



\bibitem{FronsdalInteg}
C.~Fronsdal, {\it Massless Fields with Integer Spin}, Phys. Rev. D
\textbf{18} (1978) 3624.


\bibitem{FronsdalHalfint}
J.~Fang, C.~Fronsdal, {\it Massless Fields with Half Integral Spin},
Phys. Rev. D \textbf{18} (1978) 3630.

\bibitem{Courtright} T.~Courtright, {\it Massless Field
Supermultiplets With Arbitrary Spins}, Phys. Lett. B \textbf{85}
(1979) 2019.

\bibitem{Vasiliev} M.~A.~Vasiliev, {\it Gauge form of description of
massless fields with arbitrary spin}, Yad. Fiz. \textbf{32} (1980)
855-861 (in Russian) [Sov. J. Nucl. Phys. \textbf{32} (1980) 439].

\bibitem{BKoutr} I.~L.~Buchbinder, K.~Koutrolikos, {\it BRST
Analysis of the Supersymmetric Higher Spin Models,} JHEP \textbf{12}
(2015) 106, [arXiv:1510.06569 [hep-th]].

\bibitem{Kuz1} S.~Kuzenko, A.~Sibiryakov, V.~Postnikov, {\it Massless
gauge superfields of higher half integer superspins}, JETP Lett.
\textbf{57} (1993) 534.

\bibitem{Kuz2} S.~Kuzenko, A.~Sibiryakov, {\it Massless
gauge superfields of higher integer superspins}, JETP Lett.
\textbf{57} (1993) 539.

\bibitem{Kuz3} S.~Kuzenko, A.~Sibiryakov, {\it Free massless higher
spuperspin superfields in the anti-de Sitter superspace}, Phys.
Atom. Nucl. \textbf{57} (1994) 1257, [arXiv:1112.4612 [hep-th]].

\bibitem{BKS} I.~L.~Buchbinder, S.~M.~Kuzenko, A.~G.~Sibiryakov, {\it
Quantization of higher spin superfields in the anti-de Sitter
superspace}, Phys. Lett. B \textbf{352} (1995) 29-36,
[arXiv:hep-th/9502148].

\bibitem{GKS1} S.~J.~Gates, Jr., S.~M.~Kuzenko, A.~G.~Sibiryakov,
{\it Towards unified theory of massless superfields of all
superspins,} Phys. Lett. B \textbf{394} (1997) 343,
[arXiv:hep-th/9611193].

\bibitem{GKS2} S.~J.~Gates, Jr., S.~M.~Kuzenko, A.~G.~Sibiryakov,
{\it ${\cal N}=2$ supersymmetry of higher superspin massless
theories,} Phys. Lett. B \textbf{412} (1997) 95,
[arXiv:hep-th/9609141].

\bibitem{GKuz} S.~J.~Gates, S.~M.~Kuzenko, {\it $4D, {\cal N}=1$
higher spin gauge superfields and quantized twistors,} JHEP
\textbf{0510} (2005) 008, [arXiv:hep-th/0506255].

\bibitem{Kuzenko:2017ujh}
S.~M.~Kuzenko, R.~Manvelyan, S.~Theisen, {\it Off-shell
superconformal higher spin multiplets in four dimensions,} JHEP
\textbf{07} (2017) 034,
[arXiv:1701.00682 [hep-th]].

\bibitem{HK1} J.~Hutomo, S.~M.~Kuzenko, {\it Non-conformal higher spin
supercurrents,} Phys. Lett. B \textbf{778} (2018) 242-246,
[arXiv:1710.10837 [hep-th]].

\bibitem{HK2} J.~Hutomo, S.~M.~Kuzenko, {\it The massless integer superspin
multiplets revisited,} JHEP \textbf{02} (2018) 137,
[arXiv:1711.11364 [hep-th].

\bibitem{BHK} E.~I.~Buchbinder, J.~Hutomo, S.~M.~Kuzenko, {\it Higher spin supercurrents in
anti-de Sitter space,} JHEP \textbf{09} (2018) 027,
[arXiv:1805.08055 [hep-th]].

\bibitem{Buchbinder:2019yhl}
E.~I.~Buchbinder, D.~Hutchings, J.~Hutomo, S.~M.~Kuzenko, {\it
Linearised actions for $ \mathcal{N} $ -extended (higher-spin)
superconformal gravity,} JHEP \textbf{08} (2019) 077,
[arXiv:1905.12476 [hep-th]].

\bibitem{KPR} S.~M.~Kuzenko, M.~Ponds, E.~S.~N.~Raptakis, {\it New locally (super)conformal gauge
models in Bach-flat backgrounds,} JHEP \textbf{08} (2020) 068,
[arXiv:2005.08657 [hep-th]].

\bibitem{Kuzenko:2021pqm}
S.~M.~Kuzenko, E.~S.~N.~Raptakis, {\it $\mathcal{N} = 2$
superconformal higher-spin gauge theories in four dimensions,}
[arXiv:2104.10416 [hep-th]].

\bibitem{GK1} S.~J.~Gates, Jr., K.~Koutrolikos, {\it On $4D,
{\cal N}=1$ massless gauge superfields of arbitrary superhelicity,}
JHEP \textbf{1406} (2014) 098, [arXiv:1310.7385 [hep-th]].

\bibitem{GK2} S.~J.~Gates, Jr., K.~Koutrolikos, {\it From Diophantus
to supergravity and massless higher spin multiplets,} JHEP
\textbf{1711} (2017) 063, [arXiv:1707.00194 [hep-th]].

\bibitem{BGK1} I.~L.~Buchbinder, S.~J.~Gates, K.~Koutrolikos, {\it
Superfield continuous spin eqautions of motion,} Phys. Lett. B
\textbf{793} (2019) 445-450, [arXiv:1903.08631 [hep-th]].

\bibitem{BGK2} I.~L.~Buchbinder, S.~J.~Gates, Jr., K.~Koutrolikos,
{\it Hierarchy of Supersymmetric Higher Spin Connections,} Phys.
Rev. D \textbf{102} (2020) 125018, [arXiv:20010.02061 [hep-th]].


\bibitem{Z1} Yu.~M.~Zinoviev, {\it Massive ${\cal N}=1$ supermultiplets with arbitrary
superspins,} Nucl. Phys. B \textbf{785} (2007) 98-114,
[arXiv:hep-th/0704.1535].

\bibitem{Z2} I.~L.~Buchbinder, M.~V.~Khabarov, T.~V.~Snegirev,
Yu.~M.~Zinoviev, {\it Lagrangian formulation of the massive higher
spin ${\cal N}=1$ supermultiplets in $AdS_4$ space,} Nucl. Phys. B
\textbf{942} (2019) 1-29, [arXiv:1901.09637 [hep-th]].

\bibitem{K} K.~Koutrolikos, {\it Superspace formulation
for massive half-integer superspin}, JHEP \textbf{03} (2021) 254,
[arXiv:2021.12225 [hep-th]].

\bibitem{M2} R.~R.~Metsaev, {\it Cubic interactions for arbitrary spin ${\cal N}$-extended
massless supermultiplets in $4d$ flat space,} JHEP \textbf{11}
(2019) 084, [arXiv:1909.05241 [hep-th]].

\bibitem{BS} I.~L.~Buchbinder, T.~V.~Snegirev, {\it Lagrangian formulation of free
${\cal N}$-exended massless higher spin multiplets in $4D, AdS$ space,}  Symmetry \textbf{12} (2020) 2052,
[arXiv:2009.00896 [hep-th]].

\bibitem{SeSi} A.~Yu.~Segal, A.~G.~Sibiryakov, {\it  Explicit ${\cal N}=2$ supersymmetry for higher-spin massless fields in $D=4$ AdS superspace},
Int. J. Mod. Phys. A  \textbf{17} (2002) 1207, [arXiv:hep-th/9903122].


\bibitem{M1} R.~R.~Metsaev, {\it Cubic interaction vertex for ${\cal
N}=1$ arbitrary spin massless supermultiplet in  flat space,} JHEP
\textbf{08} (2019) 130, [arXiv:1905.11357 [hep-th]].


\bibitem{KhZ}  M.~V.~Khabarov, Yu.~M.~Zinoviev, {\it Cubic interaction vertices for massless
higher spin supermultiplets in $d=4$,}  JHEP \textbf{02} (2021) 167,
[arXiv:2012.00482 [hep-th]].

\bibitem{BKTW}  I.~L.~Buchbinder, V.~A.~Krykhtin, M.~Tsulaia, D.~Weissman,
{\it Cubic Vertices for ${\cal N}=1$ Supersymmetric Massless Higher
Spin Fields in Various Dimensions,} Nucl. Phys. B \textbf{967}
(2021) 115427, [arXiv:2103.08231 [hep-th]].


\bibitem{BuchGK3} I.~L.~Buchbinder, S.~J.~Gates, K.~Koutrolikos,
{\it Conserved higher spin supercurrents for arbitrary spin massless
supermultiplets and higher spin superfield cubic interactions,} JHEP
\textbf{08} (2018) 055, [arXiv:1805.04413 [hep-th]].

\bibitem{BuchGK4} I.~L.~Buchbinder, S.~J.~Gates, K.~Koutrolikos,
{\it Integer superspin supercurrents of matter supermultiplets,}
JHEP \textbf{05} (2019) 031, [arXiv:1811.12858 [hep-th]].


\bibitem{GaK} S.~J.~Gates, K.~Koutrolikos, {\it Progress on cubic interactions of arbitrary superspin
supermultiplets via gauge invariant supercurrents,} Phys. Lett. B
\textbf{797} (2019) 134868, [arXiv:1904.13336 [hep-th]].


\bibitem{Sezgin1} E.~Sezgin, P.~Sundell, {\it Higher Spin ${\cal
N}=8$ Supergravity,}  JHEP \textbf{11} (1998) 016,
[arXiv:hep-th/9805125].

\bibitem{Engquist:2002vr}
J.~Engquist, E.~Sezgin, P.~Sundell, {\it On ${\cal N}=1, {\cal N}=2,
{\cal N}=4$ higher spin gauge theories in four-dimensions,} Class.
Quant. Grav. \textbf{19} (2002) 6175-6196,
[arXiv:hep-th/0207101 [hep-th]].

\bibitem{Sezgin2} E.~Sezgin, P.~Sundell, {\it Supersymmetric Higher Spin
Theories,} J. Phys. A \textbf{46} (2013) 214022, [arXiv:1208.6019
[hep-th]].

\bibitem{V1} M.~A.~Vasiliev, {\it Higher-spin gauge theories in four,
three and two dimensions,} Int. J. Mod. Phys. D \textbf{05} (1996)
763-797, [arXiv:hep-th/9611024].

\bibitem{V2} M.~A.~Vasiliev, {\it Higher Spin Gauge Theories: Star-Product and AdS Space,} In: ``The Many
Faces of the Superworld'', (2000) 533-610, [arXiv:hep-th/9910096].

\bibitem{V3} M.~A.~Vasiliev, {\it Higher spin gauge theories in various
dimensions,} Fortsch. Phys. \textbf{52} (2004) 702-717,
[arXiv:hep-th/0401177].

\bibitem{Galperin:1987em}
A.~S.~Galperin, N.~A.~Ky, E.~Sokatchev, {\it ${\cal N}=2$
Supergravity in Superspace: Solution to the Constraints,} Class.
Quant. Grav. \textbf{4} (1987) 1235.


\bibitem{Fradkin:1979cw}
E.~S.~Fradkin, M.~A.~Vasiliev, {\it Minimal Set of Auxiliary
Fields and S-Matrix for Extended Supergravity}, Lett. Nuovo Cim.
\textbf{25} (1979) 79 - 87; {\it Minimal set of auxiliary fields in
SO(2)-extended supergravity,} Phys. Lett. B \textbf{85} (1979)
47-51.

\bibitem{Zupnik:1998td}
B.~M.~Zupnik, {\it Background harmonic superfields in ${\cal N}=2$
supergravity,} Theor. Math. Phys. \textbf{116} (1998) 964-977,
[arXiv:hep-th/9803202].


\bibitem{Gates:1981qq}
S.~J.~Gates, Jr., W.~Siegel, {\it Linearized ${\cal N}=2$
superfield supergravity,} Nucl. Phys. B \textbf{195} (1982) 39-60.

\bibitem{Galperin:1987ek}
A.~S.~Galperin, E.~A.~Ivanov, V.~I.~Ogievetsky, E.~Sokatchev,
{\it ${\cal N}=2$ Supergravity in Superspace: Different Versions and
Matter Couplings,} Class. Quant. Grav. \textbf{4} (1987) 1255.


\bibitem{FradVas}
E.~S.~Fradkin, M.~A.~Vasiliev, {\it On the Gravitational Interaction
of Massless Higher Spin Fields}, Phys. Lett. B \textbf{189} (1987)
89-95.


\bibitem{Vasiliev:1990en}
M.~A.~Vasiliev, {\it Consistent equation for interacting gauge
fields of all spins in (3+1)-dimensions,} Phys. Lett. B \textbf{243}
(1990) 378-382.


\end{thebibliography}
\end{document}